\documentclass[12pt]{article}
\usepackage{amsmath,amsfonts,amssymb,dcolumn,lscape,graphicx,subfigure}
\usepackage{color}
\begin{document}
% Zeilenabstand (doppelt=8.2mm , ca. einfach:5.5mm)
\baselineskip=5.5mm
\bibliographystyle{aip}
\newcommand{\be} {\begin{equation}}
\newcommand{\ee} {\end{equation}}
\newcommand{\Be} {\begin{eqnarray}}
\newcommand{\Ee} {\end{eqnarray}}
\renewcommand{\thefootnote}{\fnsymbol{footnote}}
\def\a{\alpha}
\def\b{\beta}
\def\g{\gamma}
\def\G{\Gamma}
\def\d{\delta}
\def\D{\Delta}
\def\e{\epsilon}
\def\k{\kappa}
\def\l{\lambda}
\def\L{\Lambda}
\def\t{\tau}
\def\om{\omega}
\def\Om{\Omega}
\def\s{\sigma}
\def\lg{\langle}
\def\rg{\rangle}
\def\fext{{\rm f}_{\rm ext}}
\def\ga{\g_A}
\def\gb{\g_B}
\def\kAm{k_{A,m}}
\def\kAd{k_{A,d}}
\def\kBm{k_{B,m}}
\def\kBd{k_{B,d}}
\def\kXm{k_{X,m}}
\def\kXd{k_{X,d}}
\def\kA{k_A}
\def\kB{k_B}
\def\kT{k_T}
\def\KA{{\bf K}_A}
\def\KB{{\bf K}_B}
\def\KX{{\bf K}_X}
\def\KY{{\bf K}_Y}
\def\PA{{\bf P}_A}
\def\PB{{\bf P}_B}
\def\PX{{\bf P}_X}
\def\Peq{{\bf P}^{st.}}
\def\PiA{{\bf \Pi}_A}
\def\PiB{{\bf \Pi}_B}
\def\PiX{{\bf \Pi}_X}
\def\PiY{{\bf \Pi}_Y}
\def\GA{{\bf G}_A}
\def\GB{{\bf G}_B}
\def\GX{{\bf G}_X}
\def\GY{{\bf G}_Y}
\def\G{{\bf G}}
\def\hatG{\hat{\bf G}}
\def\W{{\bf W}}
\def\M{{\bf M}}
\def\V{{\bf V}}
\def\EinT{{\bf 1}^{\rm\!T}}
\newcommand{\tblue}[1]{\textcolor{blue}{#1}}
\noindent
\begin{center}
{\Large {\bf Statistics of reversible bond dynamics observed in force-clamp spectroscopy} }\\
\vspace{0.5cm}
\noindent
{\bf Gregor Diezemann$^{1,*}$, Thomas Schlesier$^1$, Burkhard Geil$^2$\\ and Andreas Janshoff$^2$} \\
{\it
$^1$Institut f\"ur Physikalische Chemie, Universit\"at Mainz,
Welderweg 11, 55099 Mainz, FRG\\
$^2$Institut f\"ur Physikalische Chemie, Universit\"at G\"ottingen,
Tammannstra\ss e 6, 37077 G\"ottingen, FRG
\\}
\end{center}
\vspace{0.5cm}
\noindent
{\it
We present a detailed analysis of two-state trajectories obtained from force-clamp spectroscopy (FCS) of reversibly bonded systems.
FCS offers the unique possibility to vary the equilibrium constant in two-state kinetics, for instance the unfolding and refolding of biomolecules, over many orders of magnitude due to the force dependency of the respective rates.
We discuss two different kinds of counting statistics, the event-counting usually employed in the statistical analysis of two-state kinetics and additionally the so-called cycle-counting.
While in the former case all transitions are counted, cycle-counting means that we focus on one type of transitions. 
This might be advantageous in particular if the equilibrium constant is much larger or much smaller than unity because in these situations the temporal resolution of the experimental setup might not allow to capture all transitions of an event-counting analysis.
We discuss how an analysis of FCS data for complex systems exhibiting dynamic disorder might be performed yielding information about the detailed force-dependence of the transition rates and about the time scale of the dynamic disorder. 
In addition, the question as to which extent the kinetic scheme can be viewed as a Markovian two-state model is discussed.
}
\section*{I. Introduction}
One of the major goals in biomolecular research is the understanding of how molecules such as nucleic acids and most prominently proteins fold into various conformations. 
Knowledge of the underlying energy landscape is needed in order to understand how the large number of possible conformations even a single chain can assume is funneled into a single stable conformation. Subtleties of the energy landscape usually remained undiscovered and consequently folding is described in the framework of a simple two-state folder as the barrier crossing is the rate limiting
step. 

With the advent of single-molecule experiments, however, the actual measurement of probability distributions came into reach that together with the tools of statistical mechanics permitted to draw a more comprehensive picture of the folding and unfolding of biomolecules. 
For instance, single-molecule FRET (F\"orster resonance energy transfer) allows the separation of folded and unfolded subpopulations, which implies that folding and unfolding can be investigated under near native conditions\cite{Schuler:2002}. 
Besides optical methods, particularly mechanical single-molecule techniques have been successfully applied to study the reversible folding and unfolding of molecules in an unprecedented 
way\cite{Junker:2009}. 
Here, addition of a denaturant is not necessary since the molecules are stretched out of their native
conformation along a predetermined coordinate.

Even the simplest version of an energy landscape, two-states separated by a single energy barrier, bears intricacies that typically escape ensemble measurements and sometimes even single-molecule approaches. 
The underlying multistate kinetics might be so involved that a single time series from, for example, photon counting statistics, is insufficient to unravel the complete mechanism. Single-molecule force spectroscopy in the reversible force-clamp mode, however, allows to deform the energy landscape by changing the externally applied force\cite{Fernandez:2004,Schlierf:2004}. 
Since the rate constants are strongly dependent on the external force, thermodynamics and kinetics can be tuned over a wide range, which essentially increases the information content that eventually might allow to access the multistate transition rates that are otherwise not amenable or at least
provide a cue whether an unfolding/folding process is Markovian.

Systems showing reversible bond-dynamics, such as certain biomolecules\cite{Manosas:2006,Liphardt:2001,Chyan:2004} or specially designed molecules\cite{Janke:2009} have been investigated experimentally by force spectroscopy in recent years and the impact of finite rebinding rates on the  resulting force-spectra has been studied theoretically\cite{Seifert:2002,Li:2006,DJ:2008}. 
While in other fields of single molecule spectroscopy (SMS)\cite{Moerner:2002,Barkai:2004} the analysis of trajectories has been used to gain information about the details of the underlying dynamical processes such an analysis has not been performed up to now in the field of FCS. 

The general theory of the treatment of the statistics of single molecule trajectories has been developed over the last decade with particular emphasis on reversible reaction 
schemes\cite{Cao:2000,GS:2006,Flomenbom:2008}. 
In particular, the analysis of the waiting time distributions (WTDs) can be used to distinguish between (non-Markovian) two-state behavior and more complex dynamics\cite{BYP:2005,FKS:2005}.
In a previous paper\cite{DJ:2009}, denoted as paper I in the following, two of us investigated the impact of non-Markovian dynamics on FCS experiments using a generating function approach introduced to the treatment of photon counting statistics by Brown\cite{Brown:2006}. 
We focussed on the mean number of transitions and its variance, the so-called Mandel parameter as a function of the external force. 
We showed, that similar to SMS, the Mandel parameter allows the detection of deviations from Markovian behavior.

In the present paper, we apply the formalism developed in the quoted papers to trajectories as  obtained from FCS experiments on systems exhibiting reversible bond-breaking dynamics. 
For example, the free energy landscape of the prominent GCN4 leucine zipper recently has been investigated in great detail as a function of external force using dual beam optical tweezers\cite{Gebhardt:2010}.
The idea of this paper is to provide a theoretical framework and feasible recepies that permit an in-depth analysis of (equilibrium) fluctuations as a function of externally applied force.
In order to do so, we consider kinetic models with two 'states' (ensembles) $A$ and $B$ (corresponding to 'bond-closed' and 'bond-open' states) and an arbitrary number of conformational substates within each of the $A$- and $B$-ensembles. 
We will show that the possibility of changing the equilibrium constant for the transitions among 
the two ensembles is a key step to determine the detailed dependence of the phenomenological kinetic rates on the external force and thus to probe various models and particularly deviations from Markovian behavior.
In favourable cases, the variation of the equilibrium constant might open the opportunity to determine the time scale of the dynamic fluctuations.

The paper is organized as follows. 
In the next section we briefly review the standard theoretical treatment of two-state trajectories and furthermore develop the techniques needed for our description of FCS.
In particular, we compare the information that can be extracted from different counting protocols.
We then propose a possible route of analyzing experimental (or simulated) trajectories and show how relevant information about the kinetics in the system investigated can be obtained.
The corresponding calculations are performed for a simple model with two conformations in each ensemble (two-configuration model or two-channel model, TCM).
\section*{II. Theory of event-counting statistics with application to FCS}
The general formalism for the analysis of two-state trajectories as observed in single-molecule experiments has been developed mainly in the past decade\cite{Cao:2000,GS:2006,SMSbook}.
In this Section, we briefly review those formal techniques that we will use to analyze the trajectories occuring in FCS of reversible systems and additionally discuss the theoretical treatment of different techniques of counting.

\begin{figure}[h!]
\centering
\vspace{-0.25cm}
\includegraphics[width=7.0cm]{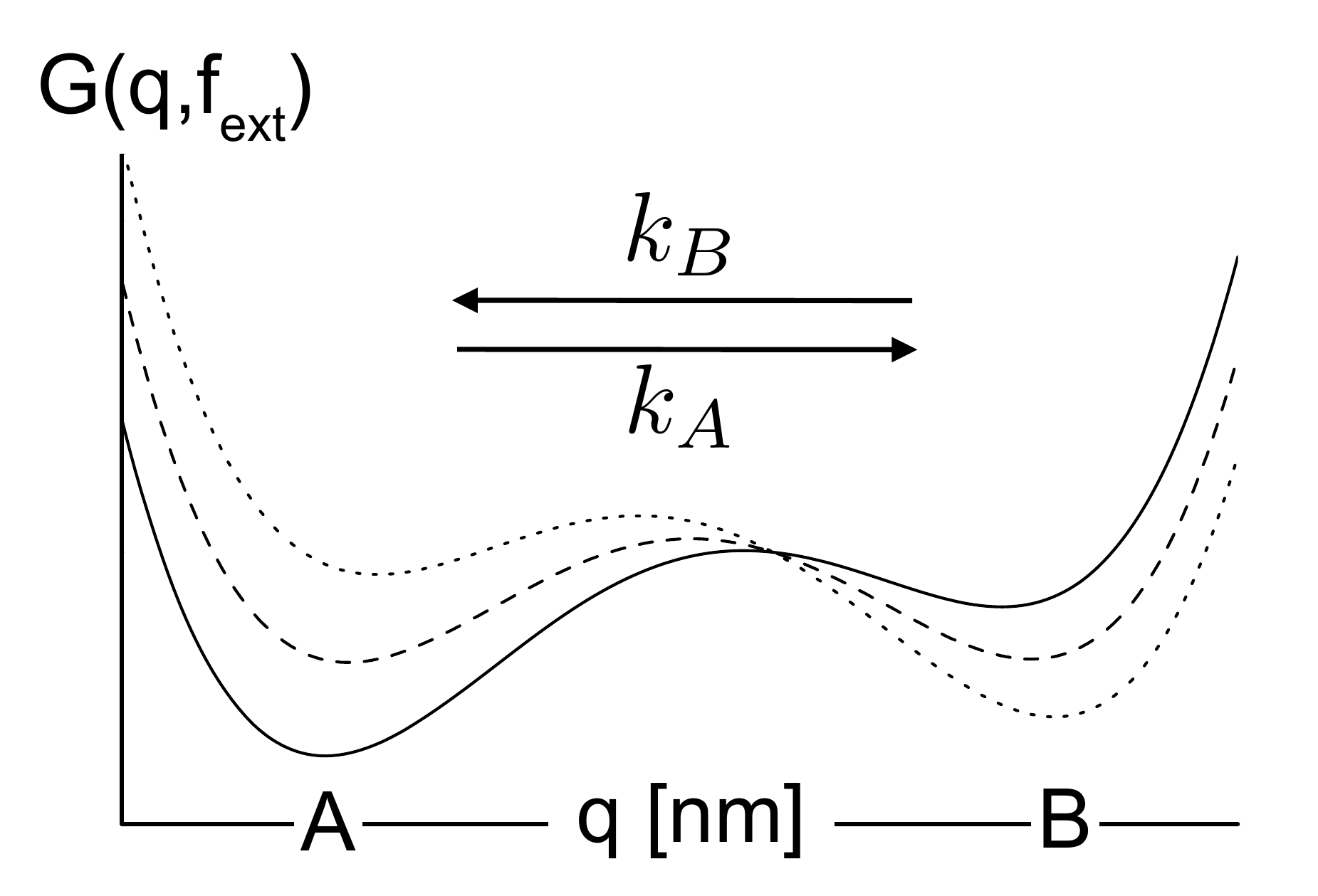}
\vspace{-0.5cm}
\caption{Sketch of the free energy $G(q,\fext)$ as a function of the reaction coordinate defined as the pulling direction for various values of the applied force.
For larger applied force, the right 'B' minimum becomes deeper (dashed and dotted lines) and the equilibrium constant $K(\fext)$ decreases.}
\label{Fig1}
\end{figure}

The simplest (coarse-grained) model of reversible bond-breaking is a two-state model for the populations of the 'states' $A$ and $B$, defined by the minimum regions in an energy landscape as sketched in Fig.\ref{Fig1}. If described as a Markov process, the master 
equation\cite{vanKampen:1981} reads:
\be\label{ME.TSM}
\partial_t{p_X(t)}=-k_X p_X(t)+k_Y p_Y(t)
\quad\mbox{with}\quad
X,Y=A,B
\quad\mbox{and}\quad
Y\neq X
\ee
where $\kA=k(A\to B)$ and $\kB=k(B\to A)$ are the corresponding transition rates.
In FCS, a constant external force $\fext$ is applied to the system under consideration (a single molecule or an adhesion cluster, for example). 
The dependence of the transition rates in an energy landscape with two (or more) minima on the external force can be approximated by Kramers theory, cf. for instance the treatment in 
ref.\cite{DJ:2008}.
Quite generally, the applied force changes the apparent activation energy and therefore one has a strong dependence of the kinetic rates on the external force, cf. Fig.\ref{Fig1}.
The activation energy for the $A\to B$-transitions will decrease as a function of the applied force $\fext$ whereas the one for the reverse transition will increase, meaning that $\kA(\fext)$ increases and $\kB(\fext)$ decreases.
Therefore, the equilibrium constant
\be\label{Kf.def}
K(\fext)=\kB(\fext)/\kA(\fext)
\ee
strongly decreases as a function of $\fext$.
As already noted in the Introduction, this fact allows to investigate vastly different physical situations by FCS that otherwise are only accessible via e.g. temperature variations in favourable cases.
In the following, we will always use the shorthand notation $K\equiv K(\fext)$ and have in mind that 
$K$ can easily be varied via variation of $\fext$. Similarly, all rates are to be understood as depending on the external force.
From the stationary solution of eq.(\ref{ME.TSM}),
\be\label{px.st.TSM}
p_A^{st.}={K\over1+K}
\quad\mbox{and}\quad
p_B^{st.}={1\over1+K}
\ee
it is evident that the system predominantly resides in one of the states if $K$ strongly deviates from unity and that the reversible nature is most pronounced for $K\sim1$.
\subsection*{A. Statistics of transitions - a brief review}
If the statistical nature of the transitions between the $A$- and $B$-states is important as is the case when dealing with trajectories, it is not sufficient to consider the global populations 
$p_X(t)$ as it is done in the master equation, eq.(\ref{ME.TSM}).
Instead, one has to treat the individual transitions and describe their statistical properties.
There are different ways to do so.
One possibility is to consider a two-component Markov process consisting of the 'reaction coordinate' and the 'number of transitions'\cite{DJ:2009,Brown:2006}.
Another way of treating the statistics is to decompose the system into two 'half-reactions' describing the decay of either $A$ or $B$ and the full kinetic scheme is then obtained by 'adding' the half-reaction schemes\cite{Cao:2000}. 
The most general procedure consists in rendering the observed transition irreversible\cite{GS:2006}.
The two ways just mentioned can be viewed as special cases of this general method.

In the following, we assume that we do not only have a single state for $A$ and $B$ but a multitude of (conformational) substates defining the $A$-ensemble and similar for the $B$-ensemble.
The populations of these ensembles are then described by the vectors 
$\PA^{\rm T}=(p_{A,1},p_{A,2},\cdots,p_{A,N_A})$ and similar for $\PB$.
The dimensions $N_X$ give the number of substates in the X-ensemble, $X=A$, $B$.
Of course, continuous models can be treated in a similar way.
Our resulting kinetic scheme then reads as:
\Be\label{ME.AB}
{d\over dt}
\left(\begin{array}{c}\PA(t)\\
\PB(t)\end{array} \right)
=\left(\begin{array}{cc}\PiA-\KA' & \KB\\
\KA & \PiB-\KB'	\end{array} \right)
\left(\begin{array}{c}\PA(t)\\
\PB(t)\end{array} \right)
\Ee
Here, the elements of the matrices $\KX$ are the inter-ensemble transition rates $k_{Y,\a;X,\a'}$ for each $(X,\a')\to(Y,\a)$ transition:
\Be\label{KX.KX'.def}
\left(\KX\right)_{\a,\a'}=&&\hspace{-0.6cm}
k_{Y,\a;X,\a'}\quad (X\neq Y)
\nonumber\\
(\KX')_{\a,\a'}=&&\hspace{-0.6cm}
\d_{\a,\a'}k_{X;\a}
\quad\mbox{with}\quad
k_{X;\a}=\sum_{\a'}k_{Y,\a';X,\a}
\Ee
Furthermore, the $\PiX$ consist of the rates for the transitions $(X,\a')\to(X,\a)$ within the $X$-ensembles, the so-called exchange rates $\g_{X;\a,\a'}$ characterizing the dynamic disorder:
\be\label{PiX.def}
\left(\PiX\right)_{\a,\a'}
=-\d_{\a,\a'}\g_{X;\a}+(1-\d_{\a,\a'})\g_{X;\a,\a'}
\quad\mbox{with}\quad
\g_{X;\a}=\sum_{\a'\neq\a}\g_{X;\a',\a}
\ee
Of course, the fact that the $\PiX$ are master operators and therefore the (conformational) dynamics within a given ensemble is Markovian can be relaxed.
The corresponding Green's function obeys:
\be\label{G.ME}
\dot\G(t)=\W\G(t)
\quad\mbox{with}\quad
\W=\left(\begin{array}{cc}\PiA-\KA' & \KB\\
\KA & \PiB-\KB'	\end{array} \right)
\ee
with the initial condition $\G(0)={\bf E}$, where ${\bf E}$ denotes the unit matrix.

In order to extract the observed transitions from the remaining unobservable ones, one decomposes the transition-rate matrix according to\cite{GS:2006}:
\be\label{W.decomp}
\W=\W'+\V
\ee
Here, $\W'$ consists of the rates for unobserved transitions and the elements of $\V$ are the rates for the observed transitions, cf. the discussion below.
Note that the system described by the transition-rate matrix $\W'$ is irreversible, i.e. there is no finite stationary population of the states.
For the Green's function of this irreversible system one has:
\be\label{Gprime}
\dot\G'(t)=\W'\G'(t)
\quad\mbox{or}\quad
\G'(t)=\exp{\left(\W't\right)}  
\ee

\begin{figure}[h!]
\centering
\vspace{-0.25cm}
\includegraphics[width=14.0cm]{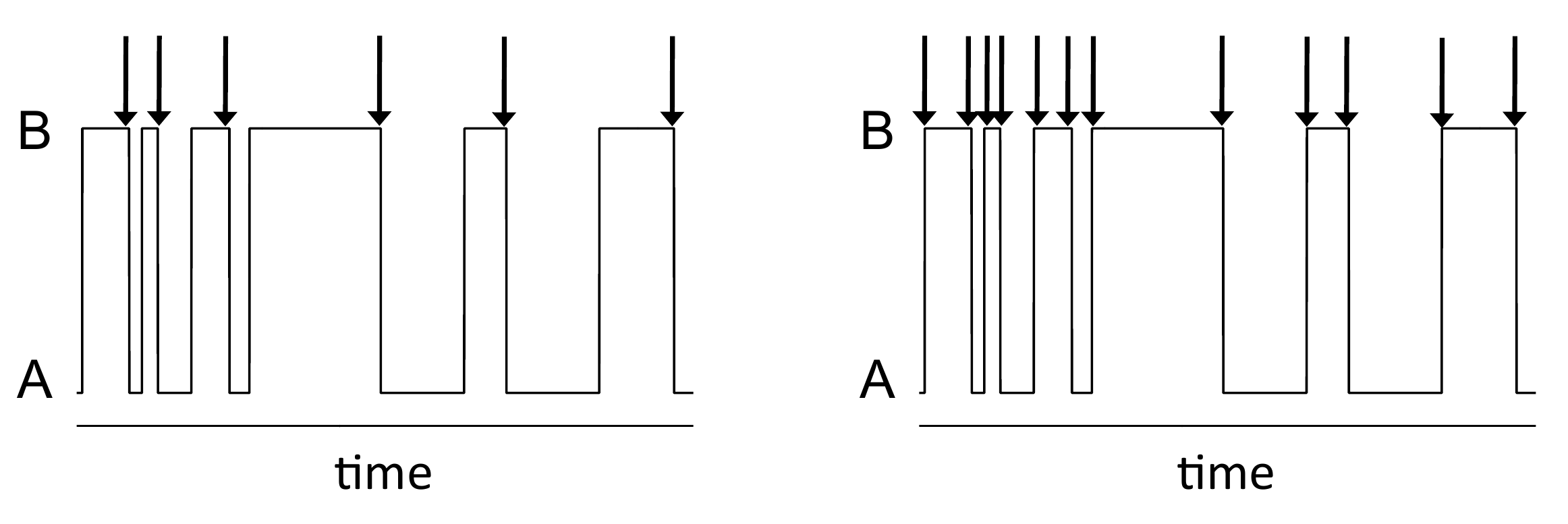}
\vspace{-0.5cm}
\caption{Example trajectories schematically showing the two different ways of counting as indicated by the arrows.
Left: cycle-counting (only $B\to A$-transitions are counted);
Right: event-counting (all transitions are counted)}
\label{Fig2}
\end{figure}

As already indicated above, in the present paper, we will consider two different choices for the observable transitions, i.e. for the matrix decomposition in eq.(\ref{W.decomp}), cf.
Fig.\ref{Fig2}.

One choice is the one that was used in our former treatment in I and consists in the consideration of whole cycles:
\be\label{V.cycle}
\W'_c=\left(\begin{array}{cc}\PiA-\KA' & 0\\
\KA & \PiB-\KB'	\end{array} \right)
\quad;\quad
\V_c=\left(\begin{array}{cc}0 & \KB\\
0 & 0	\end{array} \right)
\ee
Here, only the ($B\!\to\!A$)-transitions are counted as observable, as shown in the left part of 
Fig.\ref{Fig2}.
Of course, all results obtained from this choice are identical to the corresponding ones when interchanging $A$ and $B$.
In the following, we will denote this type of observing only full cycles as 'cycle-counting'.

The other type of decomposition is the one that typically is employed when the statistical properties of the WTDs (to be defined below) are considered.
It consists in counting all inter-ensemble transitions and will be called 'event-counting' in what follows, cf. the right part of Fig.\ref{Fig2}:
\be\label{V.events}
\W'_{ev}=\left(\begin{array}{cc}\PiA-\KA' & 0\\
0 & \PiB-\KB'	\end{array} \right)
\quad;\quad
\V_{ev}=\left(\begin{array}{cc}0 & \KB\\
\KA & 0	\end{array} \right)
\ee
Here, the underlying system is that of two half-reactions\cite{Cao:2000} and one has from 
eq.(\ref{Gprime})
\be\label{Gprime.ev}
\G'(t)_{ev}=\left(\begin{array}{cc}\GA'(t) & 0\\
0 & \GB'(t)	\end{array} \right)
\quad\mbox{with}\quad
\GX'(t)=\exp{\left\{(\PiX-\KX')t\right\}}
\ee
For this decomposition several indicators of complex dynamics have been introduced and discussed in the context of SMS\cite{Witkoskie:2004, Yang:2001, Witkoskie:2006, Cao:2006}.

An important quantity that can be extracted from single molecule trajectories and that also will be of interest for us is the distribution of the time between consecutive observable transitions\cite{GS:2006}, 
\be\label{P.t.allg.def}
{\cal P}(\t)=\EinT\V\G'(\t)\V\Peq/\lg n\rg
\ee
Here, $\EinT=(1,\cdots,1)$ is a summation row vector and the stationary populations obey $\W\Peq=0$ and are normalized, $\EinT\Peq=1$.
Furthermore, we defined the mean number of observable transitions per unit time:
\be\label{Nmit.def}
\lg n\rg=\EinT\V\Peq
\ee
Here and throughout the paper we use the stationary flux $\V\Peq/\lg n\rg$ as initial 
condition\cite{Cao:2000} because this can easily be realized by starting the counting from any event of a representative trajectory\cite{Cao:2000,GS:2006}.

In case of cycle-counting the distribution of transition times, eq.(\ref{P.t.allg.def}), corresponds to the 'turn-over time distribution':
\be\label{P.cycle.t.def}
{\cal P}_c(t)=\EinT\V_c\G_c'(t)\V_c\Peq/\lg n\rg_c
\quad\mbox{with}\quad
\lg n\rg_c=\EinT\V_c\Peq
\ee
On the other hand, for event-counting, one usually considers the WTDs\cite{Cao:2000}, given by:
\be\label{Phi.X.def}
\Phi_X(t)=\EinT_Y\KX\GX'(t)\KY\Peq_Y/\lg n\rg_X
\quad\mbox{with}\quad
\lg n\rg_X=\EinT_X\KY\Peq_Y
\quad;\quad Y\neq X =A,\,B\
\ee
Here, we defined the summation row vector $\EinT_X$ with dimension $N_X$.
In order to quantify correlations in the trajectory, one studies the distribution of consecutive waiting times:
\be\label{Phi.XY.def}
\Phi_{X,Y}(t_1,t_2)=\EinT_X\KY\GY'(t_2)\KX\GX'(t_1)\KY\Peq_Y/\lg n\rg_X
\ee
If the events are uncorrelated, one has $\Phi_{X,Y}(t_1,t_2)=\Phi_X(t_1)\Phi_Y(t_2)$ for all combinations of $X$ and $Y$ and the kinetic scheme is called reducible, meaning that it can be reduced to a two-state model, albeit usually with non-exponential WTDs\cite{FKS:2005}.

Another quantity of interest is the probability to observe a given number of transition in a prescribed time interval.
The generating function for the moments of this distribution can be written in a compact notation as\cite{GS:2006,vanKampen:1981}
\be\label{F.zt}
F(z,t)=\EinT e^{(\W'+z\V)t}\V\Peq/\lg n\rg
\ee
and the moments are computed in the standard way from $F(z,t)$ via differentiation, for instance:
\be\label{N.Nsquare}
\lg N(t)\rg=\left.{\partial F(z,t)\over\partial z}\right|_{z=1}
\quad\mbox{and}\quad
\lg N(N-1)(t)\rg=\left.{\partial^2 F(z,t)\over\partial z^2}\right|_{z=1}
\ee
from which one can calculate the Mandel parameter 
\be\label{Q.def}
Q(t)={\lg N^2(t)\rg-\lg N(t)\rg^2\over\lg N(t)\rg}-1
\ee
It can be shown that both, $\lg N(t)\rg$ and $Q(t)$, are related to certain event-correlation functions\cite{GS:2006}, but we will not dwell on this further in the present context.
We will use the expressions (\ref{N.Nsquare}) and (\ref{Q.def}) for both, event-counting and cycle-counting in the following section.
This means that we use the respective decomposition, eq.(\ref{V.cycle}) or eq.(\ref{V.events}), in the general expression for the generating function, eq.(\ref{F.zt}).
\subsection*{B. Event-counting versus cycle-counting}
As is to be expected from their definition, eq.(\ref{P.cycle.t.def}) and eq.(\ref{Phi.XY.def}),
there exists a relation between the turn-over time distribution ${\cal P}_c(t)$ and the distribution of consecutive waiting times. 
It is shown in Appendix A that this relation is given by:
\be\label{Pc.PhiAB}
{\cal P}_c(t)=\int_0^t\!d\t\Phi_{A,B}(\t,t-\t)
\ee
where we used that $\lg n\rg_c=\EinT\V_c\Peq=\EinT_A\KB\Peq_B=\lg n\rg_A$.
The system starts in state $B$ before a transition to $A$ takes place. In $A$, the system evolves for some time until it goes over to $B$ and the final $B\to A$-transition is counted. 
In the cycle-counting scheme all intermediate transition-times are summed yielding the integral in eq.(\ref{Pc.PhiAB}).
If the $A\to B$-transitions are counted as observable, one has to interchange $A$ and $B$ in the above expression.

As moments of the WTDs may be more easily extracted from experimental (or simulation) data than the 
WTDs themselves\cite{FKS:2005}, we give the expressions for these moments.
The derivation of these expressions is presented in Appendix A.
There the general expressions for the $n$th moments are given in eq.(\ref{XMoms.gen}) from which one obtains:
\be\label{Moms.PhiX}
\lg\t\rg_X=\EinT_X\Peq_X/\lg n\rg_X
\quad\mbox{;}\quad
\lg\t^2\rg_X=2\cdot\EinT_X\hatG'_X(0)\Peq_X/\lg n\rg_X
\ee
Here, $\hatG'_X(0)$ denotes $\lim_{s\to0}\hatG'_X(s)$ with 
$\hatG'_X(s)=\int_0^\infty\!dt\G'_X(t)e^{-st}$ being the Laplace transform of $\G'_X(t)$. 

We mention that the form of the exchange matrix $\PiX$ given in eq.(\ref{PiX.def}) yields
$\EinT_X\PiX=0$ and from this one can show that $\EinT_X\KY\Peq_Y=\EinT_Y\KX\Peq_X$ and thus
\be\label{nA.eq.nB}
\lg n\rg_A=\lg n\rg_B
\ee
Only in generic non-equilibrium systems we expect this relation to be violated.
The first moments of the WTDs can be used to define phenomenological rate equations.
We define the effective transition rates\cite{Yang:2001}
\be\label{kX.def}
\hat k_X=1/\lg\t\rg_X
\ee
and find, using the expression for $\lg\t\rg_X$, eq.(\ref{Moms.PhiX}), and eq.(\ref{nA.eq.nB}):
\be\label{TSM.det.balance}
\hat k_A p_A^{st.}=\hat k_B p_B^{st.}
\quad\mbox{with}\quad p_X^{st.}=\EinT_X\Peq_X
\ee
It is thus meaningful to propose the rate equation given in eq.(\ref{ME.TSM}) for the dynamical evolution of the system on a coarse-grained level, cf. the following Section.

For the moments of $\Phi_{X,Y}(t_1,t_2)$ one finds from eq.(\ref{XYMoms.gen}) 
\Be\label{Moms.PhiXY}
\lg\t\rg_{XY}=&&\hspace{-0.6cm}
\EinT_Y\hatG'_Y(0)\KX\hatG'_X(0)\Peq_X/\lg n\rg_X
\nonumber\\
\lg\t^2\rg_{XY}=&&\hspace{-0.6cm}
4\cdot\EinT_Y\left[\hatG'_Y(0)\right]^2\KX\left[\hatG'_X(0)\right]^2\Peq_X/\lg n\rg_X
\Ee
Reducible schemes can be identified via the factorization property
\be\label{Moms.XY.factor}
\lg\t^n\rg_{XY}=\lg\t^n\rg_X\lg\t^n\rg_Y
\ee

Furthermore, we can give the moments of the turn-over time distribution ${\cal P}_c(t)$.
The actual calculation is presented in Appendix A and here we give the results:
\Be\label{Moms.Pc}
\lg\t\rg_c
=&&\hspace{-0.6cm}
\lg\t\rg_A+\lg\t\rg_B
\nonumber\\
\lg\t^2\rg_c
=&&\hspace{-0.6cm}
\lg\t^2\rg_A+\lg\t^2\rg_B+2\lg\t\rg_{AB}
\Ee
The first moment in case of cycle-counting equals the sum of the moments of the WTDs relevant for event-counting and correlations between the transitions only affect the second (and higher) moments.
In particular, cycle-counting can be used in order to investigate the question whether the system under consideration can be described by reducible kinetic scheme in favourable cases.
In order to decide this, one considers the variance,
$\lg\d\t^2\rg_c=\lg\t^2\rg_c-\lg\t\rg_c^2$ which is given by
$\lg\d\t^2\rg_c=\lg\d\t^2\rg_A+\lg\d\t^2\rg_B+2(\lg\t\rg_{AB}-\lg\t\rg_A\lg\t\rg_B)$.
For a reducible scheme the last term vanishes indicating that the transitions are uncorrelated.
\section*{III. Possible analysis of trajectories}
In order to apply the theoretical results presented one has to rely on model calculations in most cases because generally valid results are rare. 
In the following we will discuss a possible way for the analysis of two-state trajectories as they can be obtained from FCS-experiments on reversibly bonded systems.
As mentioned above, one distinguishing feature of FCS is provided by the possibility of varying the 
equilibrium constant $K$ by a large amount allowing us to discuss details of the underlying kinetic scheme that are difficult to access by other means.
The particularly interesting question of deviations from Markovian/Poissonian behavior of the system under study will be treated via considering the impact of dynamic disorder\cite{Zwanzig:1990} in terms of so-called exchange models.

Before we proceed, let us briefly discuss possible origins for dynamic disorder.
One obvious reason for the existence of a large number of transition rates is given by the fact that in many biomolecules an ensemble (e.g. folded or unfolded) consists of a huge number of possible molecular conformations. Therefore, conformational changes within a given ensemble play the role of exchange processes taking place in addition to the $A\to B$-transitions.
Another possibility that has already been discussed in the literature is given by fluctuations in the transition state ensemble\cite{Raible:2006, Raible2:2006} and this has also been observed experimentally\cite{Dougan:2008}.
In terms of the simplest conceivable model, the Bell model, one has a dependence of the form
$k_{Y,\a;X,\a'}\sim e^{-\a_X\fext}$, where $\a_X$ denotes the distance to the transition state, see eq.(\ref{kX.Bell}) below\cite{Bell:1978}. 
A distribution of the latter thus results in a corresponding distribution of $k_{X,\a;Y,\a'}$. 
Sofar, the concept of these 'bond-heterogeneities' has been applied to systems showing irreversible rupture events.
An analysis of reversibly bonded (or refolding) systems opens the opportunity to study the impact of such fluctuations in the transition state ensemble in much more detail.
If the $\a_X$-fluctuations take place on the experimental time scale, the system exhibits the features of dynamic disorder.

In the present paper, we will mainly use the TCM with only two substates in each ensemble for illustrative calculations but we try to keep the  discussion as general as possible.
Details of the calculations concerning the TCM are presented in Appendix C for convenience of the reader.
Before we start considering the impact of dynamic disorder, we propose to extract averaged information from the trajectories.\\
\subsection*{A. Determination of force-dependent mean kinetic rates}
We suggest that the first step in the analysis of two-state trajectories consists in the determination of the first moments of the WTDs. 
This should be possible with high statistical significance and allows to define the effective transitions rates $\hat k_X$ on a phenomenological level, $k_X=\lg\t\rg_X^{-1}$ cf. eq.(\ref{kX.def})\cite{Yang:2001}.
(We will simply write $k_X$ instead of $\hat k_X$ in the remainder of this section.)
The resulting phenomenological two-state model consists of only a single $A$- and a single $B$-state and the temporal evolution of the populations of these states is governed by the master equation, 
eq.(\ref{ME.TSM}).

\begin{figure}[h!]
\centering
\vspace{-0.25cm}
\includegraphics[width=14.0cm]{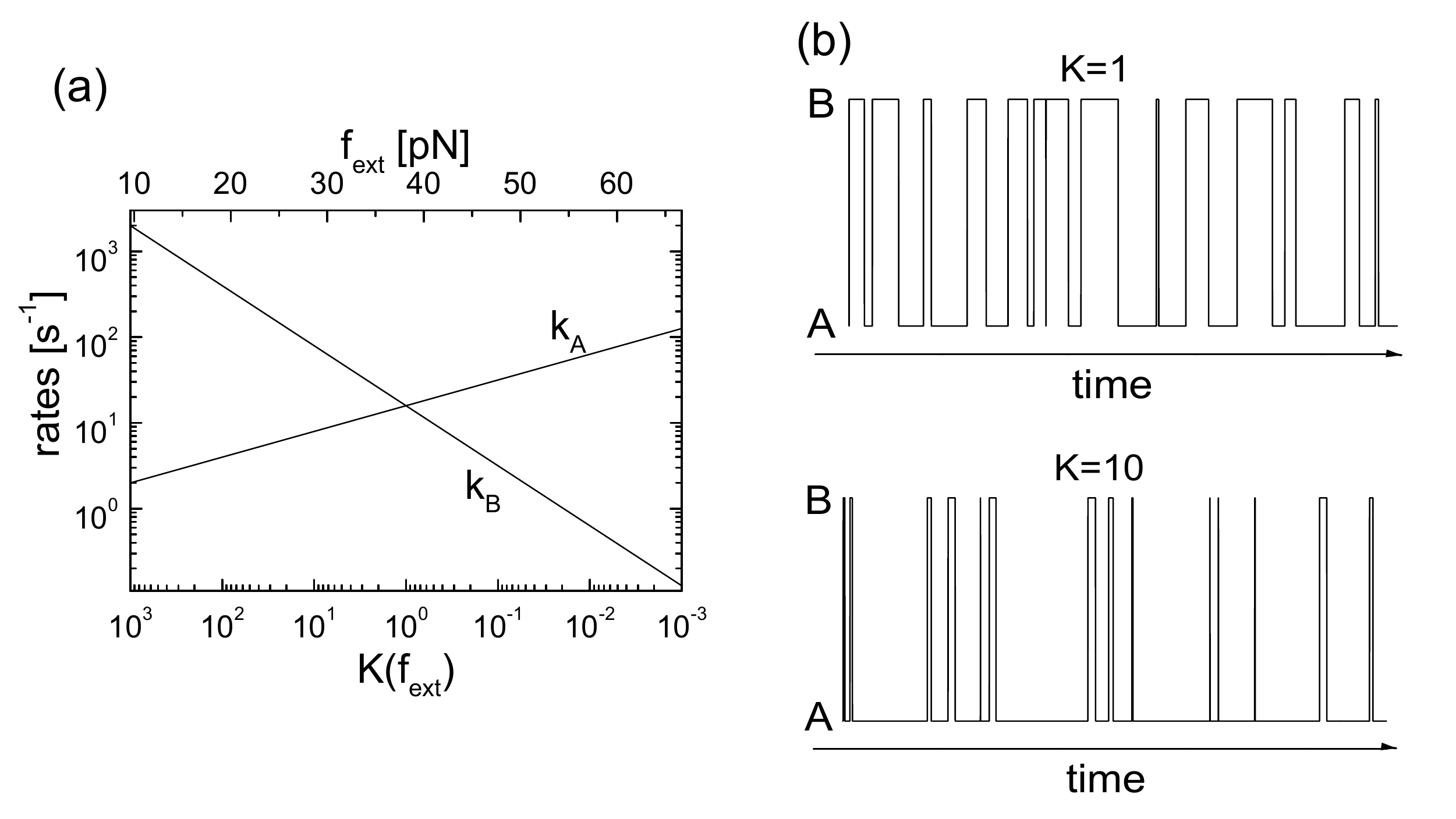}
\vspace{-0.5cm}
\caption{(a): kinetic rates $\kA=\lg\t\rg_A^{-1}$ and $\kB=\lg\t\rg_B^{-1}$ as a function of the external force (top) or of the equilibrium constant $K(\fext)$ (bottom).
The parameters are chosen as $\a_A/\b=0.3nm$ and $\a_B/\b=0.7nm$ and the temperature is 
$\b^{-1}=300K=4.14 pNnm$. Furthermore, $\kA(0)=1s^{-1}$ and $\kB(0)=10^4\kA(0)$.
(b): Typical trajectories of the two-state model for $K=1$ and $K=10$ obtained from kinetic Monte Carlo simulations for the same parameters.
}
\label{Fig3}
\end{figure}

As pointed out several times, a broad range of kinetic rates $k_X(\fext)$ is accessible via FCS.
This is because already in the phenomenological Bell model\cite{Bell:1978} one has an exponential
dependence of the kinetic rates on the external force,
\be\label{kX.Bell}
\kA(\fext)=\kA(0)e^{\a_A\fext}
\quad\mbox{and}\quad
\kB(\fext)=\kB(0)e^{-\a_B\fext}
\ee
where $\a_X$ are the distances from the transition state location scaled by the inverse temperature
$\b=1/T$ (we set the Boltzmann constant to unity). 
There are thus two parameters that are to be determined from the $\lg\t\rg_X$ as a function of the external force.
Note that in the Bell model one neglects any $\fext$-dependence of the positions of the free energy minima and the transition state and also higher order powers of 
$\fext$\cite{Manosas:2006,DJ:2009,Dudko:2006,Dudko:2008}.
In the present paper, we will use the simple form given in eq.(\ref{kX.Bell}) for model calculations and all times will be given in units of $\kA(0)$, cf. Fig.\ref{Fig3}(a). 
Because all of our arguments also apply to other single-molecule techniques and are not restricted to an analysis of FCS-data, in the following we will always refer to the equilibrium constant instead of the external force.
Of course, the simple form for the rates given in eq.(\ref{kX.Bell}) can be relaxed but at the expense of introducing more parameters. 

Experimentally, it should be possible to determine the $\fext$-dependence of the kinetic rates for a rather large range of forces. 
In particular, we propose that it is possible to quantify deviations from the simple Bell model without relying on some specific model for the $\fext$-dependence. 
This fact distinguishes FCS from standard applications of force spectroscopy using a linear force ramp\cite{Evans:2007}, where a model-free analysis appears to be difficult\cite{Dudko:2006}.

In this context, the importance of the opportunity to apply different counting schemes becomes apparent.
When considering trajectories with different values of the equilibrium constant, as shown in 
Fig.\ref{Fig3}(b) for $K=1$ and $K=10$, it is obvious that for a finite temporal resolution an analysis in terms of event-counting becomes difficult for situations in which $K$ differs strongly from unity.
In such cases, cycle-counting appears better suited and can also be used to determine $\lg\t\rg_X$ as for $K\gg1$ one has $\lg\t\rg_c=\lg\t\rg_A$ and for $K\ll1$ $\lg\t\rg_c=\lg\t\rg_B$ holds, cf.
eq.(\ref{Moms.Pc}).
Furthermore, for values of $K$ that allow an analysis in terms of both counting schemes a check for 
internal consistency of the data analysis is given by the general relation 
$\lg\t\rg_c=\lg\t\rg_A+\lg\t\rg_B$.

An independent test of the analysis performed can be obtained via a comparison of the values for $k_X$ determined from the first moments of the WTDs with the mean number of transitions in a given time 
$t$, $\lg N(t)\rg$.
The calculation of the full time-dependent expressions for $\lg N(t)\rg^{(ev)}$ and 
$\lg N(t)\rg^{(c)}$ are outlined in Appendix B and the results are given in eq.(\ref{N.ev.c.TSM}).
From these expressions one can see that in both cases after a time on the order of the inverse relaxation rate $(\kA+\kB)$ a behavior linear in time is reached:
\be\label{N.ev.c.long.TSM}
\lg N(t)\rg^{(ev)}=2\lg N(t)\rg^{(c)}=\lg n\rg t
\quad;\quad t\gg(\kA+\kB)^{-1}
\ee
Here, the mean number of observable transitions per unit time according to 
eq.(\ref{Nmit.def}) is ($\lg n\rg^{(ev)}=\lg n\rg_A+\lg n\rg_B$):
\be\label{N.mit.TSM}
\lg n\rg\equiv\lg n\rg^{(ev)}=2\lg n\rg^{(c)}={2\kA\kB\over(\kA+\kB)}
\ee
As is to be expected, in case of event-counting one has twice the number of transitions as compared to cycle-counting.

From the above analysis it should be possible to obtain the detailed dependence of the phenomenological kinetic rates on the external force.
Therefore, one can analyse this force-dependence in more detail and for instance use existing models in order to extract information about the location of transition states in the free energy landscape and details of their properties.

\begin{figure}[h!]
\centering
\vspace{-0.25cm}
\includegraphics[width=7.5cm]{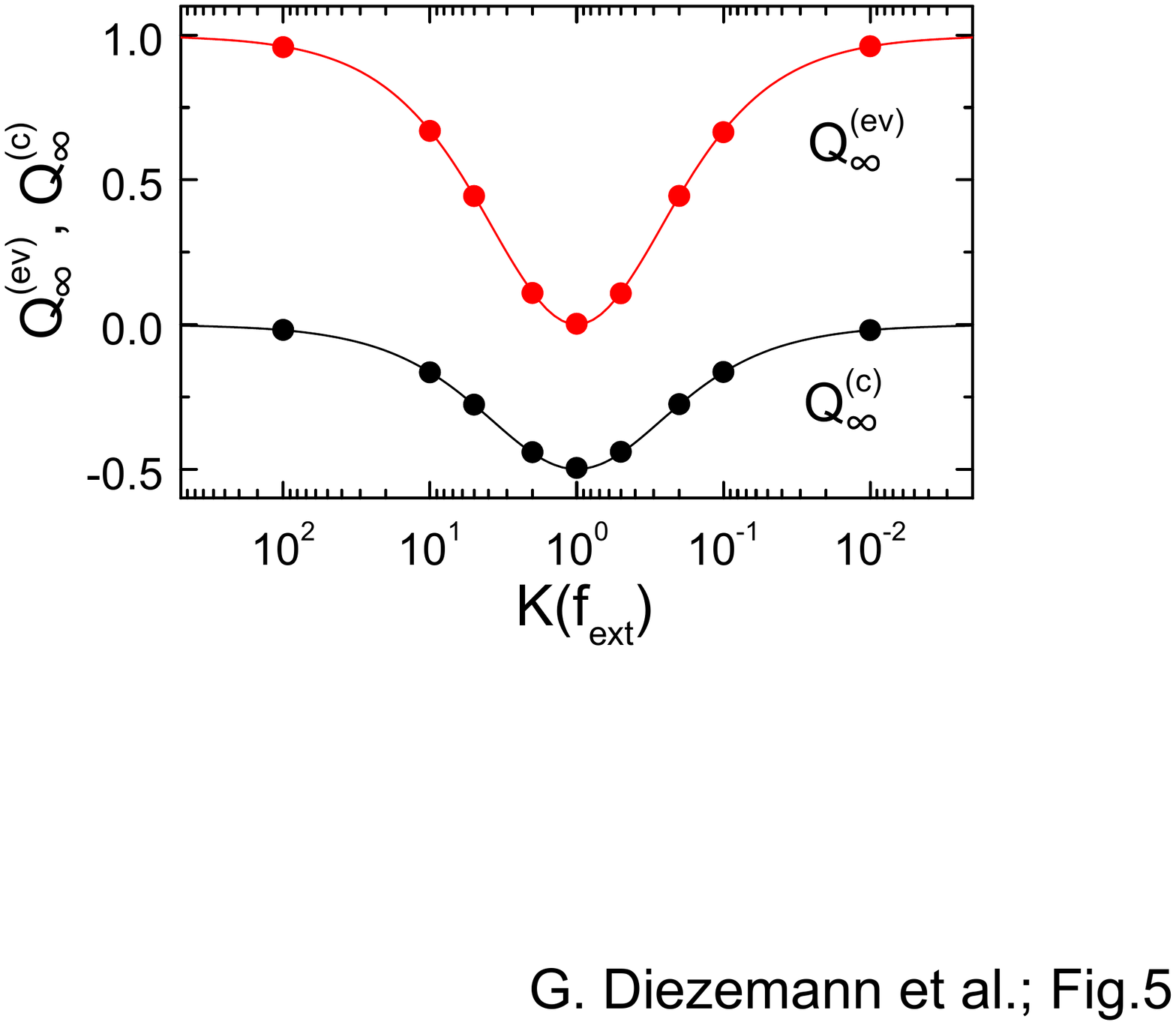}
\vspace{-0.5cm}
\caption{$Q_\infty^{(ev)}$ (red, upper line) and $Q_\infty^{(c)}$ (black, lower line) as a function of the equilibrium constant.
The parameters are the same as in Fig.\ref{Fig3}. The full lines are the analytical expressions,
eq.(\ref{Q.infty.ev.c.TSM}), and the points are results of a kinetic Monte Carlo simulation.}
\label{Fig4}
\end{figure}

As a second step in the analysis of two-state trajectories we propose to check for deviations from simple Markovian two-state behavior of the kinetic scheme under study.
As already discussed in I, the Mandel parameter is well suited to detect deviations from Markovian behavior.
The Q-parameters $Q(t)^{(ev)}$ and $Q(t)^{(c)}$ are time-dependent on a scale of $(\kA+\kB)$ and then approach their constant long-time limits $Q^{(ev)}_\infty$ and $Q^{(c)}_\infty$. 
The calculation closely follows the corresponding one of $\lg N(t)\rg$ and one finds, using the definition eq.(\ref{Q.def}):
\be\label{Q.infty.ev.c.TSM}
Q^{(ev)}_\infty={(K-1)^2\over(K+1)^2}
\quad\mbox{and}\quad
Q^{(c)}_\infty=-2{K\over(K+1)^2}
\ee
In Fig.\ref{Fig4}, we show $Q_\infty^{(ev)}$ and $Q_\infty^{(c)}$ as a function of the equilibrium constant and thus of the external force, cf. Fig.\ref{Fig3}.
By definition of the Mandel parameter, one has deviations from Poissonian behavior whenever 
$Q\neq 0$.
It is shown in Appendix B that in case of event-counting one has Poissonian behavior for $K=1$
and for cycle-counting one finds a Poisson distribution for $K\ll1$ or $K\gg1$.
Because for the determination of the moments of the event number probability one has to count all relevant transitions, we expect that also for the determination of the  Q-parameters cycle-counting will become the more reliable method for $K$ strongly deviating from unity.
The information content, however, is the same for both counting schemes.

Non-Markovian effects will of course also be reflected in the WTDs. 
In particular, deviations from exponentiality indicate complex kinetic behavior. 
As has been noted by Flomenbom et al.\cite{FKS:2005}, experimentally it usually will be easier to determine the moments of the WTDs quite accurate whereas the the WTDs will often be very noisy.
As shown in Appendix B, for the phenomenological two-state model, the WTDs are exponential distributions, $\Phi_X(t) = k_X e^{-k_X t}$, and the corresponding moments are trivially related,
$\lg\t^2\rg_X=2\lg\t\rg_X^2$.
Furthermore, the kinetic scheme is reducible, $\Phi_{X,Y}(t_1,t_2)=\Phi_X(t_1)\cdot\Phi_Y(t_2)$.

If the Q-parameter as a function of the equilibrium constant follows eq.(\ref{Q.infty.ev.c.TSM}) and the second moment of the WTDs obey $\lg\t^2\rg_X=2\lg\t\rg_X^2$, one can safely conclude that the system under study can be described by a simple kinetic two-state model and the analysis is complete.
\subsection*{B. Analysis of complex kinetic schemes}
If the above analysis of the two-state trajectories indicates deviations from Markovian behavior no general procedure of quantifying these can be proposed.
This is because the analysis of two-state trajectories does not allow to determine the underlying kinetic scheme unambiguously\cite{Flomenbom:2008,BYP:2005,FKS:2005}. 
The impact of dynamic disorder or dynamic heterogeneities can thus only be discussed in terms of specific models and different indicators aiming at quantifying the relevant time scale have been 
introduced\cite{Witkoskie:2004, Yang:2001, Witkoskie:2006, Cao:2006}.
When treating the effect of conformational changes or dynamic disorder via the introduction of exchange rates $\g_{X;\a,\a'}$ in eq.(\ref{PiX.def}), one can distinguish two limiting dynamical regimes.
For exchange rates large compared to the inter-ensemble transition rates, the so-called motional narrowing regime, one obtains the phenomenological two-state model just discussed.
In the other limiting case of vanishing $\g_{X;\a,\a'}$ the relaxation properties are determined completely by the distribution of the $k_{X,\a;Y,\a'}$ (static disorder regime).
In both cases, the time scale of the dynamic disorder cannot be determined from the data.

If deviations from Markovian two-state behavior are observed, there are two questions to be posed.
First, one would like to know the time scale of the dynamic disorder.
Furthermore, it is interesting to determine whether the system still is reducible meaning that one has two-state behavior.
This latter point is of particular interest when dealing with biomolecules like proteins, where two-state behavior is not always quantified easily.
\subsubsection*{Time scale of fluctuations}
One limiting scenario that should clearly be observable in the data is the case in which the time-dependent Q-parameter does not reach a constant limit $Q_\infty$ at long times. 
As this limit is reached on the time scale of the slowest rate in the system, one can conclude that 
on the experimental time scale one is in the static disorder limit.
A further analysis of the data for the Mandel parameter usually will be difficult because numerical calculations show that the linear $t$-dependence of $Q(t)$ at long times as given in 
eq.(\ref{Q.gamto0.TCM}) for the TCM will only be reached on a time scale much longer than the inverse inter-ensemble rates.

In this limit of static disorder, one can give general expressions for the moments of the WTDs.
Defining the average of quantities in the $X$-ensemble via
\be\label{AX.mitX}
\lg A\rg_X=\sum_{\a=1}^{N_X}A_\a p_{X,\a}^{st.}
\ee
one finds for instance
\[
\lg n\rg_X=\EinT_X\KY\Peq_Y
=\sum_{\a=1}^{N_X}k_{X;\a}p_{X,\a}^{st.}=\lg k_X\rg_X
\]
where $k_{X;\a}$ denotes the overall escape rate from the substate ($X,\a$), 
cf. eq.(\ref{KX.KX'.def}).
The first moment of the WTDs follows from eq.(\ref{kX.def}):
\be\label{tauX}
\lg\t\rg_X=\hat k_X^{-1}=p_X^{st.}/\lg k_X\rg_X
\ee
with $p_X^{st.}=\EinT_X\Peq_X$, cf. eq.(\ref{TSM.det.balance}).
For the second moment, $\lg\t^2\rg_X$, one can give an analytical expression because in the static disorder limit one has $[\hatG'_X(0)]_{\a,\k}=\d_{\a,\k}k_{X;\a}^{-1}$.
Using this expression, one finds from eq.(\ref{Moms.PhiX}) that
\be\label{tauX.h2.fd}
\lg\t^2\rg_X^{\rm{(s.d.)}}=2\lg k_X\rg^{-1}_X\left\lg k_X^{-1}\right\rg_X
\ee
The analysis of the mean rates discussed above yields information about $\lg k_X\rg_X$ and one can try to obtain the width of the underlying distribution of rates from $\lg k_X^{-1}\rg_X$.
For instance, for the TCM, one has $\lg k_X^n\rg_X=p_X^{st.}\lg k_X^n\rg$ with $p_X^{st.}$ given in eq.(\ref{px.st.TSM}) and one finds $\lg k_X^{-1}\rg=\lg k_X\rg/[\lg k_X\rg^2-\lg\d k_X^2\rg]$.

\begin{figure}[h!]
\centering
\vspace{-0.25cm}
\includegraphics[width=7.5cm]{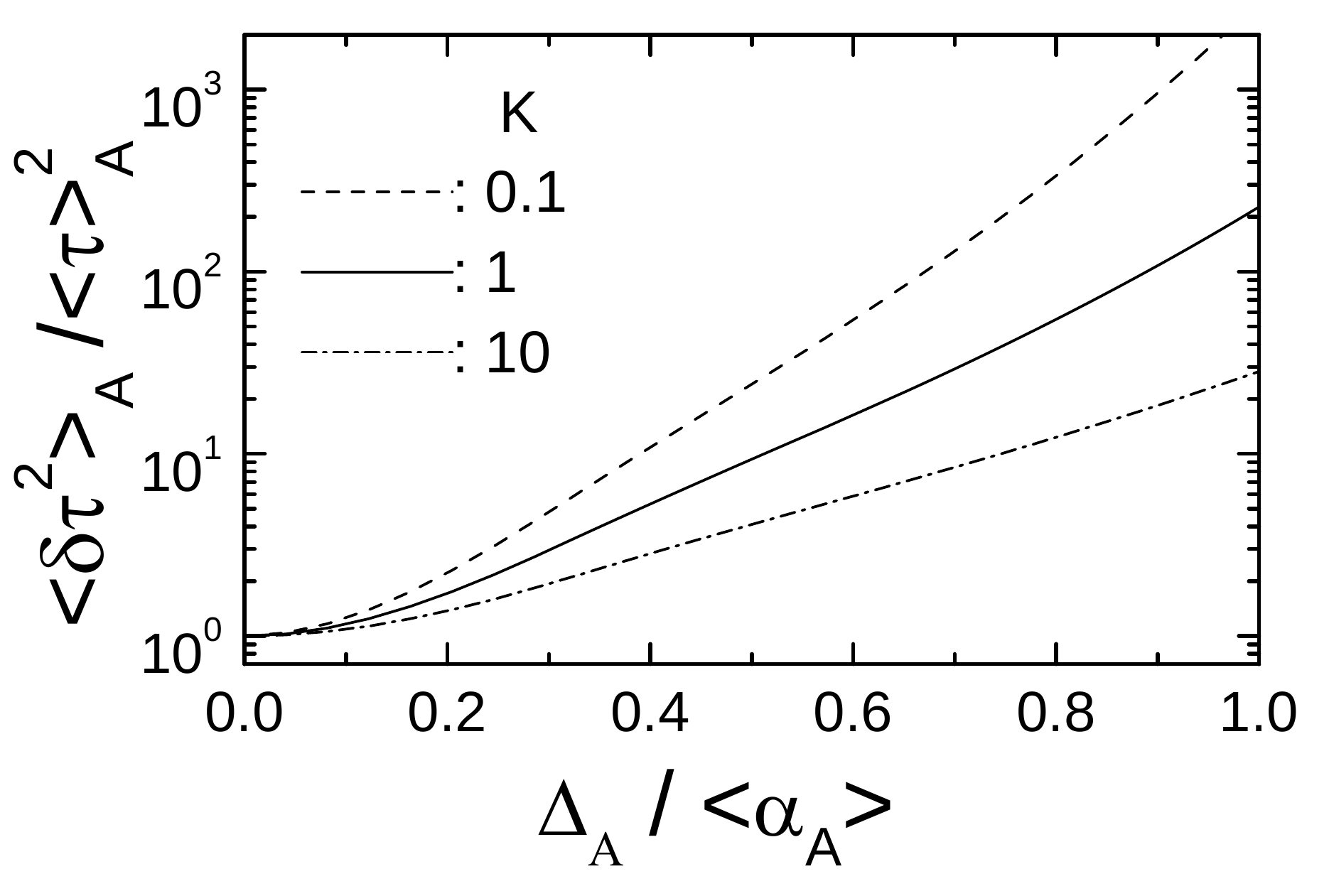}
\vspace{-0.5cm}
\caption{
$\lg\d\t^2\rg_A/\lg\t\rg_A^2$ for a model of bond-heterogeneities in the static disorder 
limit\cite{Raible:2006, Raible2:2006} as a function of the width $\D_A$ of the Gaussian distribution of the distance to the transition state, 
$g(\a_A)=N^{-1}\exp{[-(\a_A-\lg\a_A\rg)^2/(2\D_A^2)]}\Theta(\a_A)$.
Here, we used $\lg\a_A\rg/\b=0.3nm$.
}
\label{Fig5}
\end{figure}

Another example that can be used to illustrate the relation between the different moments is the model introduced by Raible et al. mentioned above\cite{Raible:2006, Raible2:2006}. 
Using the Bell expression given in eq.(\ref{kX.Bell}), $k_A(\a_A,\fext)=k_A(0)e^{\a_A\fext}$ and a Gaussian distribution of the form $g(\a_A)=N^{-1}\exp{[-(\a_A-\lg\a_A\rg)^2/(2\D_A^2)]}\Theta(\a_A)$
one finds that $\lg k_A^n\rg_X=p_X^{st.}\int_0^\infty\!d\a_A g(\a_A)k_A(\a_A,\fext)^n$.
The deviations from Markovian behavior are best characterized in terms of the ratio
$\lg\d\t^2\rg_X/\lg\t\rg_X^2=\lg\t^2\rg_X/\lg\t\rg_X^2-1$ which equals unity in the Markovian limit. 
In the static disorder limit relevant here one finds 
\[
\lg\d\t^2\rg_X/\lg\t\rg_X^2=2\lg k_X\rg_X\left\lg k_X^{-1}\right\rg_X-1
\]
In Fig.\ref{Fig5} we show this quantity as a function of the width of the $\a_A$-distribution.
It is obvious, that independent of the equilibrium constant the Markovian limit is reached for small widths $\D_A$.
For larger widths a change in the equilibrium constant could help to give an estimate for the width of the distribution.
(When applied to experimental data, values $\D_A\sim(0.5\cdots1)\lg\a_A\rg$ have been 
found\cite{Raible2:2006}.)

\begin{figure}[h!]
\centering
\vspace{-0.25cm}
\includegraphics[width=7.5cm]{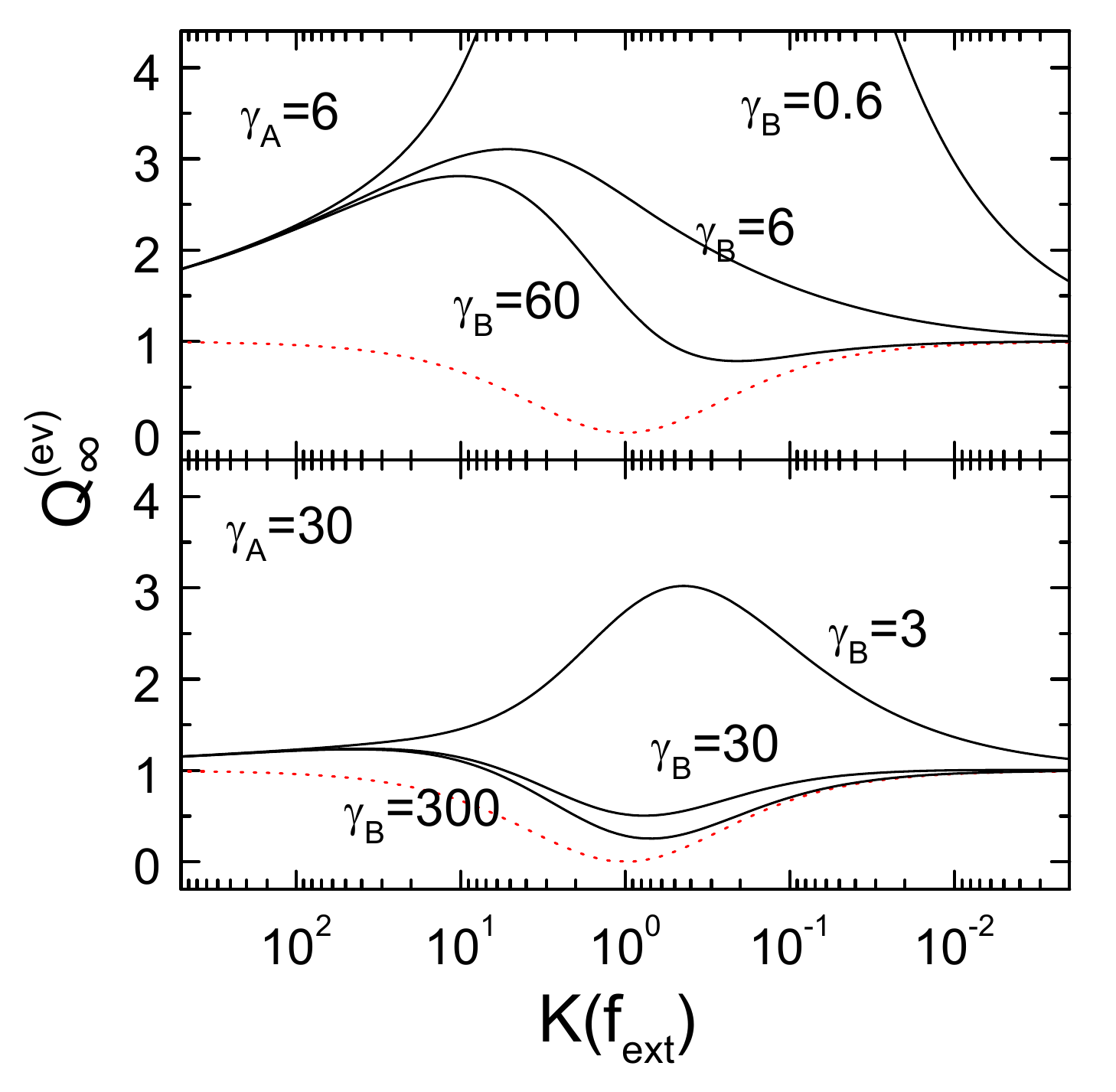}
\vspace{-0.5cm}
\caption{$Q_\infty^{(ev)}$ versus external force for various exchange rates $\ga$ and $\gb$ for the TCM defined in Appendix C.
The parameters chosen are $k_{A,a}(\fext)=1.99\bar k(\fext)$, $k_{A,b}(\fext)=0.01\bar k(\fext)$, where $\bar k(\fext)=\kA(0)e^{\a_A\fext}$ as defined in eq.(\ref{kX.Bell}).
This means that according to eq.(\ref{kXM.kXD.def}) one has $\lg k_X\rg=\bar k$ and 
$\s_X=0.995\bar k$.
The other parameters are those given in the caption to Fig.\ref{Fig3}.
The dotted red line is the Markovian limit given in eq.(\ref{Q.infty.ev.c.TSM}).}
\label{Fig6}
\end{figure}

A possible cross-over from the static disorder limit to dynamic disorder on the time scale of the experiment/simulation is most easily observed when considering $Q(t)$.
If the time scale of the dynamic disorder is reached in the time window used in the analysis, 
$Q(t)$ is known to reach a finite $t$-independent value at long times.
This limit $Q_\infty^{(ev)}$ is shown in Fig.\ref{Fig6} for the TCM for various values of the exchange rates $\ga$ and $\gb$. 
It appears that it is mainly the smaller exchange rate that determines the overall behavior and also the obvious deviations from the Markovian value (dotted lines).
As we have already noted in I, the determination of the Mandel parameter allows the 'detection' of deviations from Markovian behavior but this does not allow to discriminate among different kinetic schemes.
However, a variation of the equilibrium constant might be used to attempt a determination of the mean exchange rate.

\begin{figure}[h!]
\centering
\vspace{-0.25cm}
\includegraphics[width=7.5cm]{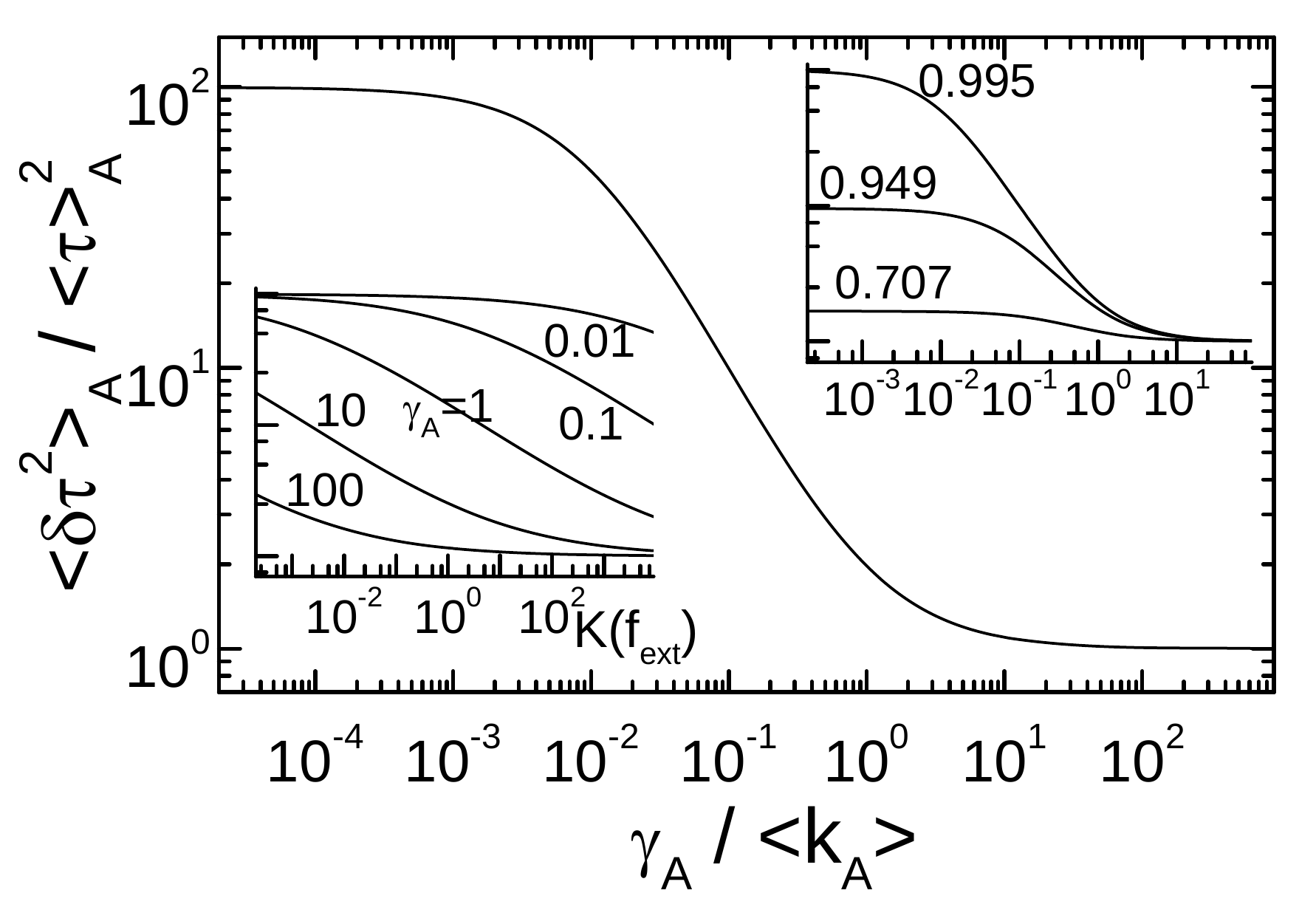}
\vspace{-0.5cm}
\caption{$\lg\d\t^2\rg_A/\lg\t\rg_A^2$ versus $\g_A/\lg k_A\rg$ for the TCM showing the transition from the static disorder limit for small $\g_A$ to the motional narrowing regime for large $\g_A$.
The upper right inset shows the dependence on the 'width of the rate distribution', $\s_A$.
Here, the values used are 
$0.995$: ($k_{A,a}(\fext)=1.99\bar k(\fext)$, $k_{A,b}(\fext)=0.01\bar k(\fext)$);
$0.949$: ($k_{A,a}(\fext)=1.9\bar k(\fext)$, $k_{A,b}(\fext)=0.1\bar k(\fext)$);
$0.707$: ($k_{A,a}(\fext)=1.5\bar k(\fext)$, $k_{A,b}(\fext)=0.5\bar k(\fext)$).
The left inset shows $\lg\d\t^2\rg_A/\lg\t\rg_A^2$ as a function of the equilibrium constant indicating that in some cases the cross-over between different regimes might be observable.
The remaining parameters are the same as in Fig.\ref{Fig6}.
}
\label{Fig7}
\end{figure}

We have pointed out above that deviations from Markovian two-state behavior also manifest themselves in the deviations of the moments of the WTDs from the simple behavior $\lg\t^2\rg_X=2\lg\t\rg_X^2$.
In Fig.\ref{Fig7} we show $\lg\d\t^2\rg_A/\lg\t\rg_A^2$ for the TCM as a function of the exchange rate. 
(The expressions for the relevant moments are given in eq.(\ref{tau.n.x.TCM}).)
In the static disorder limit one has the expression given above and in the limit of motional narrowing, the model reduces to the Markovian two-state model and one has 
$\lg\d\t^2\rg_A=\lg\t\rg_A^2$.
The dependence on the width parameter is shown in the inset. Also shown is the dependence on the external force in terms of the equilibrium constant. From this plot it becomes apparent that in favourable cases one should be able to observe a transition from static to dynamic disorder or from dynamic disorder to motional narrowing.
For the specific model considered here, we find that one has about three to four orders of magnitude of exchange rates which define the regime of dynamic disorder.
In this regime, one might attempt to determine the exchange rates using one of the methods that have been discussed by Cao and coworkers\cite{Witkoskie:2004, Yang:2001, Witkoskie:2006, Cao:2006}.
Additionally, the long-time limit of the Q-parameter can be helpful to determine the mean time scale of the fluctuations.
\subsubsection*{Reducibility of the kinetic scheme}
We have mentioned already that the question whether the kinetic scheme under study can be considered as a two-state system is of interest in many applications, in particular when concerned with the folding dynamics of biomolecules.
As discussed in the context of eq.(\ref{Phi.XY.def}), a kinetic scheme is reducible, if the joint probability function $\Phi_{X,Y}(t_1,t_2)$ of successive waiting times factorizes, 
$\Phi_{X,Y}(t_1,t_2)=\Phi_X(t_1)\Phi_Y(t_2)$, for all combinations $X$, $Y$.
We note that $\Phi_{X,Y}(t_1,t_2)$ contains all information about correlations in the trajectory and in principle is obtained from the data by constructing a histogram of the intersections of successive 
$(X\!\to\!Y)$-transitions followed by $(Y\!\to\!X)$-transitions\cite{Flomenbom:2008}.
In ref.\cite{FKS:2005} the various connectivities giving rise to a reducible scheme are discussed. 
In Appendix A, we show that in the static disorder limit also models with a global connectivity among the substates of the two ensembles can be reducible for a certain form of the transition rates, cf. eq.(\ref{kYX.product}).

In order to test experimental data for reducibility, in general one has to perform a detailed comparison of the WTDs $\Phi_A(t)$, $\Phi_B(t)$ and $\Phi_{X,Y}(t_1,t_2)$ or the corresponding moments.
In particular, the determination of $\Phi_{X,Y}(t_1,t_2)$ (or the corresponding moments 
$\lg\t^n\rg_{XY}$) from event-counting statistics might be difficult with the required accuracy. 
In this case, cycle-counting can provide the desired information with much higher significance.
A comparison of the variances $\lg\d\t^2\rg_c$ and $\lg\d\t^2\rg_A$, $\lg\d\t^2\rg_B$
yields important information about $\lg\t\rg_{AB}$, cf. the discussion at the end of Section II.

\begin{figure}[h!]
\centering
\vspace{-0.25cm}
\includegraphics[width=7.5cm]{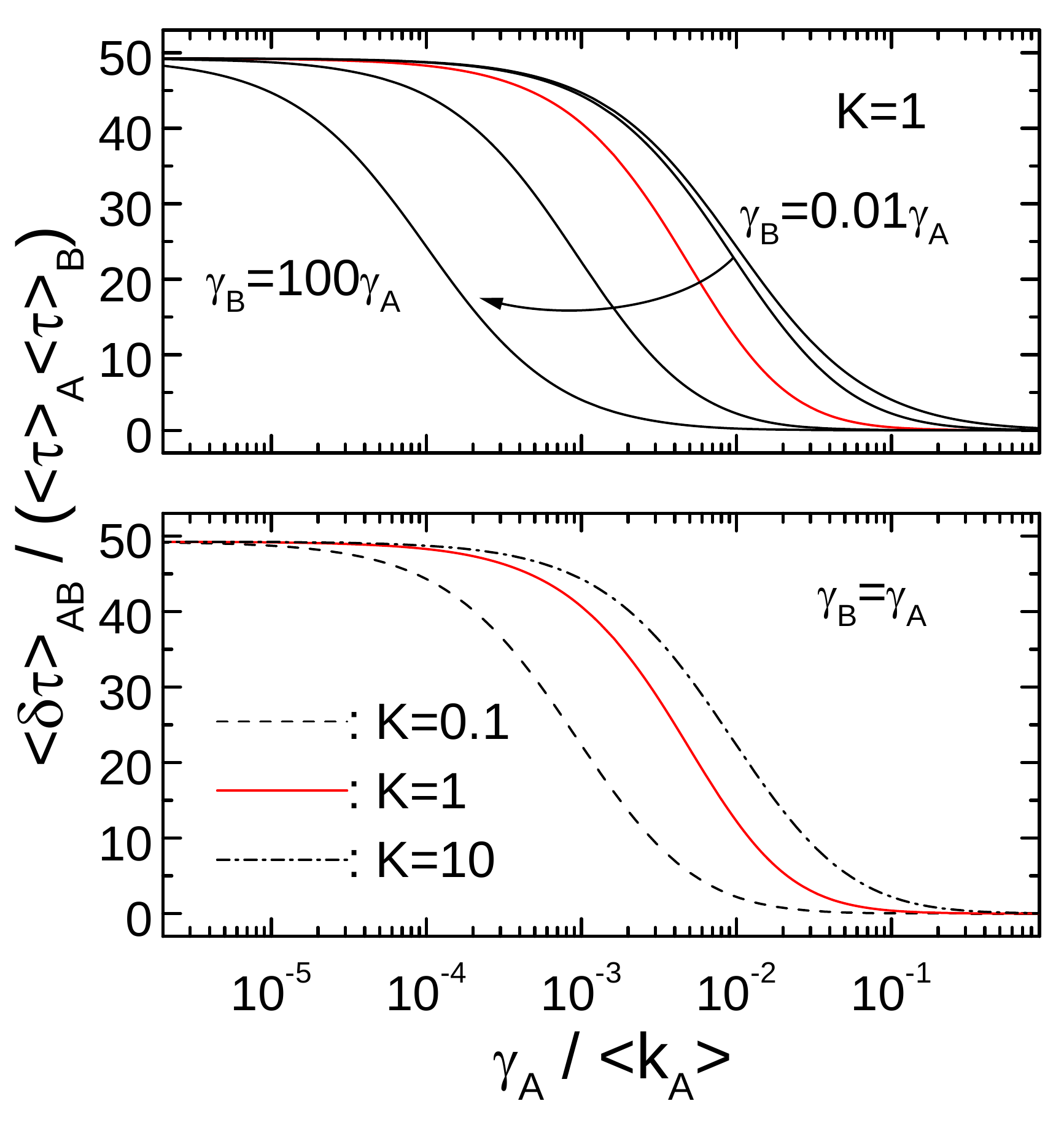}
\vspace{-0.5cm}
\caption{$\lg\d\t\rg_{AB}/(\lg\t\rg_A\lg\t\rg_B)$ versus $\g_A/\lg k_A\rg$ for the TCM.
Upper panel: different exchange rates are assumed for the $A$- and the $B$-ensemble,
$\g_B/\g_A=0.01,0.1,1,10,100$ from right to left. 
The transition to the reducible regime depends on their ratio.
The red line is for $\g_B=\g_A$.
Lower panel: transition to reducibility as a function of the equilibrium constant.
The parameters are the same as in Fig.\ref{Fig6}.
}
\label{Fig8}
\end{figure}

We show the behavior of $\lg\d\t\rg_{AB}=\lg\t\rg_{AB}-\lg\t\rg_A\lg\t\rg_B$ for the TCM as a function of the exchange rate in Fig.\ref{Fig8}, cf. eq.(\ref{del.tau.n.XY.res.einK}).
It is evident that the system becomes reducible for large exchange rates, i.e. the motional narrowing regime.
The transition to this reducible regime depends on the ratio of the exchange rates in the two ensembles (upper panel) and also on the equilibrium constant (lower panel).
In some favourable situations it might be possible to observe the transition to the reducible regime via variation of the external force. 
It is obvious from the expression for $\lg\d\t\rg_{AB}$ that the maximum value increases with larger $\s_X$. 
Apart from the motional narrowing limit, the TCM becomes reducible in the limit $k_{A,b}\to0$ in which case the kinetic scheme consists of a single 'gateway state'\cite{FKS:2005}.

Another situation resulting in a reducible scheme that might be of experimental relevance is given by the limit of one exchange rate $\g_X$ being much larger than all other rates, cf. Fig.\ref{Fig8} (upper panel).
This scenario might apply if the $X$-ensemble represents a stiff bonded system (with many closed hydrogen bonds) and in the other ensemble the bonds are opened and therefore there is more structural  flexibility for conformational fluctuations.
The dynamics of the escape from the $X$-ensemble is a Poisson process but the dynamic disorder present in $Y(\neq X)$ is reflected in the measurable quantities like $Q_\infty^{(ev)}$ (cf. 
Fig.\ref{Fig6}).
\section*{V. Conclusions}
We have investigated the statistics of two-state trajectories as they are expected to occur in force-clamp spectroscopy of reversibly bonded systems.
As has been shown earlier, one can detect the deviations from Markovian behavior from an analysis of the moments of the distribution of the overall number of events.
In particular, the Mandel parameter allows the direct observation of dynamic disorder.
We have discussed two different ways of analyzing the trajectories. 
One can either count all transitions, denoted as event-counting, or one counts only one kind of transitions (($A\to B$) or ($B\to A$)), termed cycle-counting. 
We have shown the interrelation between the WTDs and their first two moments for the two counting schemes.
Depending on the value of the equilibrium constant one of the methods might be superior to the other if applied to real data with finite temporal resolution.
In particular, for values of the equilibrium constant strongly deviating from unity, the system mainly resides in one of the states (either $A$ or $B$) and event-counting might not be feasible.
In this situation, cycle-counting still gives access to most dynamical information required to analyse the data.

We propose that the first moments of the WTDs as a function of the external force can be used for an  analysis of the dependence of the kinetic rates on the external force. 
In particular, one does not rely on fitting explicit expressions derived from some model to the data.
The consisteny of the analysis can be checked by comparing the phenomenological transition rates obtained from the first moments of the WTDs with the mean number of events.

An analysis beyond the determination of the mean rates can be proposed and allows to determine whether the kinetic scheme is determined by dynamic disorder or if one is in one of the limiting situations of static disorder or motional narrowing.
In order to discuss these issues we have performed model calculations for one particular kinetic scheme, a two-configuration (two-channel) model, but we expect the general features not to depend too sensitively on the model used.
In particular, the Mandel parameter should behave quite different in the mentioned limits.
Thus, its determination gives a first hint on the time scale of dynamic fluctuations.

As the various moments depend on the inter-ensembles transition rates and the exchange rates in different ways, we expect that a comparison of the first and second moments of the WTDs as a function of the force dependent equilibrium constant can be helpful in the determination of the average exchange rate. 
This holds in particular, if the indicators of dynamic disorder that have been discussed in the past are applied additionally\cite{Witkoskie:2004, Yang:2001, Witkoskie:2006, Cao:2006}.
Also the long-time limit of the Mandel parameter can be used for the purpose of determining the fluctuation time scale.
It is a major advantage of FCS to provide a means of varying the equilibrium constant by a huge amount due to its dependence on the external force.
Therefore, FCS might be superior to other methods in determining the life-time of the dynamic disorder.

Even though a distinction of different kinetic schemes is not possible in general, consideration of the moments of the WTDs for event-counting and cycle-counting might in favourable cases allow to determine whether the kinetic scheme underlying the observed two-state dynamics is reducible or not.
This analysis might be very helpful in the discussion of the folding pathways in biomolecules.

In summary, we propose to analyze two-state trajectories as they can be determined from FCS experiments in order to gain information about the force-dependence of the kinetic rates, the deviations from Markovian behavior and about the time scale of dynamic disorder, if present. 
We hope that our analysis will be helpful in the design of explorative experiments in the future.
\section*{Acknowledgment}
We thank Thorsten Metzroth, Gerald Hinze and J\"urgen Gauss for fruitful discussions.
Financial support by the Deutsche Forschungsgemeinschaft via the SFB 625 is acknowledged.
\begin{appendix}
\section*{Appendix A: Some properties of the WTDs}
\setcounter{equation}{0}
\renewcommand{\theequation}{A.\arabic{equation}}
In this Appendix, we derive the relation between the WTDs and the turn-over times distribution and give the general expressions for the moments of the WTDs.
\subsection*{Relation between ${\cal P}_c(t)$ and $\Phi_{X,Y}(t_1,t_2)$}
We start from the Laplace transform of eq.(\ref{P.cycle.t.def}),
\be\label{LT.Pc}
\hat{\cal P}_c(s)=\EinT\V_c\G_c'(s)\V_c\Peq/\lg n\rg_c
\ee
and write, using the definition for $\W_c'$, eq.(\ref{V.cycle})
\[
\G_c'(s)=
\left(\begin{array}{cc}s{\bf E}_A-\W_A' & 0\\
-\KA & s{\bf E}_B-\W_B'	\end{array}\right)^{-1}
\]
where we used the abbreviation:
\be\label{WX.prime}
\W_X'=\PiX-\KX'
\ee
Recognizing that $\GX'(s)^{-1}=(s{\bf E}_X-\W_X')$ and using the formula for the inversion of a block matrix, one finds:
\[
\G_c'(s)=
\left(\begin{array}{cc}\GA'(s) & 0\\
\GB'(s)\KA\GA'(s) & \GB'(s)\end{array} \right)
\]
Inserting this decomposition into eq.(\ref{LT.Pc}) yields
\be\label{LT.Pc.PhiAB}
\hat{\cal P}_c(s)=\EinT_A\KB\GB'(s)\KA\GA'(s)\KB\Peq_B/\lg n\rg_A
\ee
because of $\lg n\rg_c=\EinT\V_c\Peq=\lg n\rg_A$.
If after inverse Laplace transform this expression is compared to the definition of 
$\Phi_{A,B}(t_1,t_2)$ in eq.(\ref{Phi.XY.def}) one obtains eq.(\ref{Pc.PhiAB}).

\subsection*{Moments of the WTDS}
The moments of $\Phi_X(t)$ are defined in the following way:
\be\label{tau.def}
\lg\t^n\rg_X
=\int_0^\infty\!dt t^n\Phi_X(t)
\quad\mbox{and}\quad
\lg\t^n\rg_{XY}
=\int_0^\infty\!dt_1\int_0^\infty\!dt_2 t_1^nt_2^n\Phi_{X,Y}(t_1,t_2)
\ee
We start from the definition of the WTD, eq.(\ref{Phi.X.def}), and use the following identities:
\be\label{Ident1}
\EinT_X\PiX=0
\quad\mbox{and}\quad
\EinT_X\KY{\bf A}_Y=\EinT_Y\KY'{\bf A}_Y
\ee
where ${\bf A}_Y$ is an arbitrary vector of dimension $N_Y$.
We will furthermore use the detailed balance condition
\be\label{det.balance}
\W\Peq=0
\;\to\;
\W_X'\Peq_X=-\KY\Peq_Y
\ee
where $\Peq=(\Peq_A,\Peq_B)^{\rm T}$ and the second identity follows from the structure of the transition rate matrix $\W_{ev}'$ given in eq.(\ref{V.events}) and additionally we used the definition of $\W_X'$, eq.(\ref{WX.prime}).
We now use these expressions to slightly rewrite eq.(\ref{Phi.X.def}) for $\Phi_X(t)$.
In the expression $\EinT_Y\KX\GX'(t)\KY\Peq_Y$, we use eq.(\ref{Ident1}) in the form
$\EinT_Y\KX{\bf A}_X=\EinT_X\KX'{\bf A}_X=-\EinT_X(\PiX-\KX'){\bf A}_X=-\EinT_X\W_X'{\bf A}_X$ and eq.(\ref{det.balance}) to get:
\be\label{Phi.alternativ}
\Phi_X(t)=\EinT_X\W_X'\GX'(t)\W_X'\Peq_X/\lg n\rg_X
\ee
Using the definition of the Laplace transform, ${\hat f}(s)=\int_0^\infty\!dtf(t)e^{-st}$, one finds for the moments:
\be\label{Xmoms.LT}
\lg\t^n\rg_X=(-1)^n\lim_{s\to0}
\EinT_X\W_X'{d^n\over ds^n}\hatG_X'(s)\W_X'\Peq_X/\lg n\rg_X
\ee
There are different ways to obtain expressions for the derivatives needed here.
One way is to start from the equation for $\hatG_X'(s)$, 
$s\hatG_X'(s)-{\bf E}_X=\W_X'\hatG_X'(s)$, with ${\bf E}_X$ denoting the $N_X^2$-dimensional unit matrix, to multiply this by powers of $s$ and to differentiate the result.
This is explained in detail in the Appendices of ref.\cite{GS:2006}.
This procedure holds for arbitrary transition rate matrices,
However, in our case, the transition rate matrix $\W_X'$ has a special form because it is the rate matrix for an irreversible system.
In particular, we can safely assume that there is no eigenvalue $\l=0$.
This is because the matrix $\PiX$ has exactly one vanishing eigenvalue\cite{vanKampen:1981}, corresponding to the stationary solution of the problem and $\KX'$ consists of at least one non-vanishing matrixelement.
Thus, the matrix $\W_X'$ has no vanishing eigenvalues and the inverse $(\W_X')^{-1}$ exists.
We can thus use the form 
\be\label{GX.Series}
\hatG_X'(s)=\left[s{\bf E}_X-\W_X'\right]^{-1}
={\hatG_X'(0)\over{\bf E}_X+s\hatG_X'(0)}
=\hatG_X'(0)\sum_{n=0}^\infty s^n \left[-\hatG_X'(0)\right]^n
\ee
with $\hatG_X'(0)=-(\W_X')^{-1}$.
From this series, one derives
\be\label{GX.derivative}
(-1)^n\lim_{s\to0}{d^n\over ds^n}\hatG_X'(s)=n!\left[\hatG_X'(0)\right]^{n+1}
\ee
This yields, using eq.(\ref{Xmoms.LT}):
\be\label{XMoms.gen}
\lg\t^n\rg_X=n!
\EinT_X\left[\hatG_X'(0)\right]^{n-1}\Peq_X/\lg n\rg_X
\ee
Similarly, we obtain for the moments of the distribution of consecutive waiting times, again using
eq.(\ref{Ident1}) and eq.(\ref{det.balance}) along with the two-dimensional Laplace transform
${\hat f}(s_1,s_2)=\int_0^\infty\!dt_1\int_0^\infty\!dt_2f(t_1,t_2)e^{-s_1t_1}e^{-s_2t_2}$:
\be\label{XYmoms.LT}
\lg\t^n\rg_{XY}=\lim_{s_2\to0}\lim_{s_1\to0}
\EinT_Y\W_Y'{d^n\over ds_2^n}\hatG_Y'(s_2)\KX{d^n\over ds_1^n}\hatG_X'(s_1)\W_X'\Peq_X/\lg n\rg_X
\ee
This expression yields with the aid of eq.(\ref{GX.derivative}):
\be\label{XYMoms.gen}
\lg\t^n\rg_{XY}=(n!)^2
\EinT_Y\left[\hatG_Y'(0)\right]^n\KX\left[\hatG_X'(0)\right]^n\Peq_X/\lg n\rg_X
\ee
From these general equations, eq.(\ref{Moms.PhiX}) and eq.(\ref{Moms.PhiXY}) given in the text are derived.

The results for the moments of the turn-over time distribution given in eq.(\ref{Moms.Pc}) are obtained in the following way.
One uses the definition $\lg\t^n\rg_c=(-1)^n\lim_{s\to0}{d^n\over ds^n}\hat{\cal P}_c(s)$ with
$\hat{\cal P}_c(s)$ given in eq.(\ref{LT.Pc.PhiAB}).
This expression is rearranged into the form 
\[
\hat{\cal P}_c(s)=\EinT_B\W_B'\GB'(s)\KA\GA'(s)\W_A'\Peq_A/\lg n\rg_A
\]
by applying the identities (\ref{Ident1}) and (\ref{det.balance}).
Using eq.(\ref{GX.derivative}) for the derivatives, after some algebra one obtains eq.(\ref{Moms.Pc}). 
\subsection*{WTDs in the frozen disorder limit}
The limit of vanishing exchange rates $\g_{X;\a,\a'}$ can be treated analytically because in this case one has 
\be\label{GX.fd}
\left[\GX'(t)\right]_{\a,\a'}=\d_{\a,\a'}e^{-k_{X;\a}t}
\quad;\quad\g_{X;\a,\a'}\to0 
\ee
where $k_{X;\a}=\sum_{\a'}k_{Y,\a';X,\a}$, cf. eq(\ref{KX.KX'.def}).
Using this expression, the WTDs are calculated according to eq.(\ref{Phi.X.def}) and 
eq.(\ref{Phi.XY.def}) with the result:
\Be\label{WTDs.fd}
\Phi_X(t)=&&\hspace{-0.6cm}
\lg k_X\rg_X^{-1}\sum_\a \left(k_{X;\a}\right)^2e^{-k_{X;\a}t}p_{X,\a}^{st.}
\nonumber\\
\Phi_{X,Y}(t_1,t_2)=&&\hspace{-0.6cm}
\lg k_X\rg_X^{-1}\sum_{\a,\a'}k_{Y;\a}e^{-k_{Y;\a}t_2}k_{Y,\a;X,\a'}
e^{-k_{X;\a'}t_1}k_{X;\a'}p_{X,\a'}^{st.}
\Ee
Furthermore, the moments of $\Phi_X(t)$ are given by eq.(\ref{tauX}) and eq.(\ref{tauX.h2.fd}) and those of $\Phi_{X,Y}(t_1,t_2)$ are found to be given by:
\be\label{tauXYhn.fd}
\lg\t^n\rg_{XY}=(n!)\lg k_X\rg_X^{-1}
\sum_{\a,\a'}k_{Y;\a}^{-n}k_{Y,\a;X,\a'}k_{X;\a'}^{-n}p_{X,\a'}^{st.}
\ee
Eq.(\ref{WTDs.fd}) shows that the kinetic scheme is reducible if the rate constants are chosen according to
\be\label{kYX.product}
k_{Y,\a;X,\a'}=\lg k_X\rg_X^{-1}\cdot p_{Y,\a}^{st.}k_{Y;\a}\cdot k_{X;\a'}
\ee
and the indices $\a$ and $\a'$ are independent of each other.
\section*{Appendix B: The phenomenological two-state model}
\setcounter{equation}{0}
\renewcommand{\theequation}{B.\arabic{equation}}
In this Appendix we give details of the calculations for some expressions relevant for the analysis of trajectories.
\subsubsection*{Generating functions}
In order to compute $F(z,t)$, one uses eq.(\ref{F.zt}) and chooses the appropriate expression for 
$\V$, i.e. $\V_{c}$ for cycle-counting and $\V_{ev}$ in case of event-counting.
For a two-state model, the corresponding matrices directly follow from eq.(\ref{V.cycle}) and 
eq.(\ref{V.events}) and are explicitly given by:
\[
\W_{c}'=\left(\begin{array}{cc}-k_A & 0\\
k_A & -k_B	\end{array} \right)
\,\mbox{;}\,
\V_{c}=\left(\begin{array}{cc}0 & k_B\\
0 & 0	\end{array} \right)
\]
and
\[
\W_{ev}'=\left(\begin{array}{cc}-k_A & 0\\
0 & -k_B	\end{array} \right)
\,\mbox{;}\,
\V_{ev}=\left(\begin{array}{cc}0 & k_B\\
k_A & 0	\end{array} \right)
\]
In the expression for the generating function, $F(z,t)=\EinT e^{(\W'+z\V)t}\V\Peq/\lg n\rg$, one furthermore needs the equilibrium populations given in eq.(\ref{px.st.TSM}).
As all matrices are two-dimensional, the diagonalization can be performed analytically and the matrix-exponential $e^{(\W'+z\V)t}$ can be computed.
This way, one finds in case of event-counting:
\be\label{F.ev.zt}
F_{ev}(z,t)={1\over2w_{ev}}
\left[
w_{ev}\left(e^{-k_{ev}^{(-)}t}+e^{-k_{ev}^{(+)}t}\right)
+z(1+K)\left(e^{-k_{ev}^{(-)}t}-e^{-k_{ev}^{(+)}t}\right)
\right]
\ee
where
\[
w_{ev}=\sqrt{(1-K)^2+4Kz^2}
\quad\mbox{and}\quad
k_{ev}^{(\pm)}={\kA\over2}\left(1+K\pm w_{ev}\right)
\]

\noindent
Similarly, for cycle-counting one has:
\be\label{F.c.zt}
F_{c}(z,t)={1\over2w_{c}}
\left[
w_{c}\left(e^{-k_{c}^{(-)}t}+e^{-k_{c}^{(+)}t}\right)
+(1+K)\left(e^{-k_{c}^{(-)}t}-e^{-k_{c}^{(+)}t}\right)
\right]
\ee
with
\[
w_{c}=\sqrt{(1-K)^2+4Kz}
\quad\mbox{and}\quad
k_{c}^{(\pm)}={\kA\over2}\left(1+K\pm w_{c}\right)
\]
\subsubsection*{Statistics of transitions}
From the generating functions the mean number of transitions and the Mandel parameter are calculated as derivatives of the generating function with respect to $z$ as given in eq.(\ref{N.Nsquare}) and one finds
\Be\label{N.ev.c.TSM}
\lg N(t)\rg^{(ev)}=&&\hspace{-0.6cm}
{(\kA-\kB)^2\over 2(\kA+\kB)^2}\left(1-e^{-(\kA+\kB) t}\right)+\lg n\rg t
\nonumber\\
\lg N(t)\rg^{(c)}=&&\hspace{-0.6cm}
{\lg n\rg\over 2(\kA+\kB)}\left(e^{-(\kA+\kB) t}-1\right)+{1\over2}\lg n\rg t
\Ee
where $\lg n\rg=2k_Ak_B/(k_A+k_B)$, cf. eq.(\ref{N.mit.TSM}).
From these expressions it becomes evident that the linear long-time behavior given in eq.(\ref{N.ev.c.long.TSM}) is reached after a time on the order of $(\kA+\kB)^{-1}$.
Also the Q-parameters $Q(t)^{(ev)}$ and $Q(t)^{(c)}$ are time-dependent on the same time scale and then approach their constant long-time limits $Q^{(ev)}_\infty$ and $Q^{(c)}_\infty$. 
The calculation closely follows the corresponding one of $\lg N(t)\rg$, but the resulting expressions are not very impressive. 
The values obtained for the long-time limits are given in the text in eq.(\ref{Q.infty.ev.c.TSM}).

In order to show in which limiting case Poisson statistics is observed, one has to recognize that this statistics is characterized by an exponentially decaying generating function of the form
$F(z,t)=\exp{(-(1-z)\k t)}$ with some rate $\k$, yielding $((\k t)^n/n!)e^{-\k t}$ for the corresponding probability\cite{GS:2006}.

\noindent
It is easy to show that for event-counting one has 
\be\label{F.ev.poisson}
K=1:\quad F_{ev}(t)=e^{-(1-z)\kA t}=e^{-(1-z)\kB t}
\ee
and in case of cycle-counting one finds
\be\label{F.c.poisson}
K\ll1:\quad F_{c}(t)=e^{-(1-z)\kB t}
\quad\mbox{and}\quad
K\gg1:\quad F_{c}(t)=e^{-(1-z)\kA t}
\ee
\subsubsection*{Waiting time distributions}
The WTDs are calculated according to eq.(\ref{Phi.X.def}) using the fact that here the matrices 
$\KX$ are one-dimensional and consist only of the rates $k_X$ and the stationary populations of the states are given in eq.(\ref{px.st.TSM}).
As expected, the resulting WTDs are exponential distributions, $\Phi_X(t)=k_X e^{-k_Xt}$, and the corresponding moments are given by $\lg\t^n\rg_X=n!k_X^{-n}$ and in particular $\lg\t\rg_X=k_X^{-1}$, cf. eq.(\ref{kX.def}).
Furthermore, $\Phi_{X,Y}(t_1,t_2)$ is obtained from eq.(\ref{Phi.XY.def}) and one finds that 
\[
\Phi_{X,Y}(t_1,t_2)=\Phi_X(t_1)\cdot\Phi_Y(t_2)
\]
which means that this simple kinetic scheme is reducible.
\section*{Appendix C: The TCM}
\setcounter{equation}{0}
\renewcommand{\theequation}{C.\arabic{equation}}
\subsubsection*{Definition of the model}
\begin{figure}[h!]
\centering
\vspace{-0.25cm}
\includegraphics[width=7.0cm]{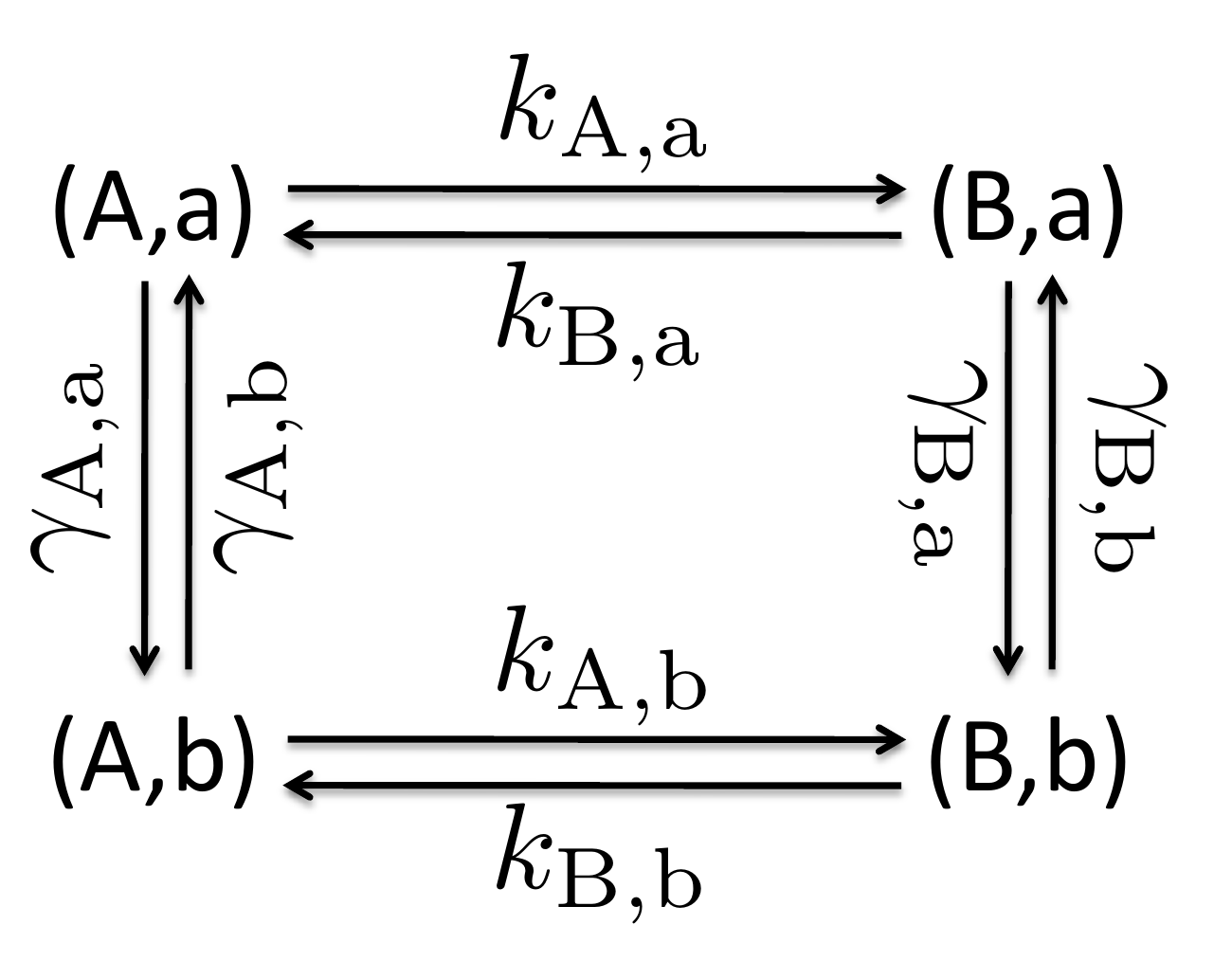}
\vspace{-0.5cm}
\caption{Definition of the transition rates in the TCM.
We abbreviate the exchange rate $\g_{X;b}=\g_{X;a,b}$ and $\g_{X;a}=\g_{X;b,a}$ for $X=A$, $B$.
Similarly, $k_{X,a}$ and $k_{X,b}$ are abbreviations for $k_{Y,b;X,a}$ and $k_{Y,a;X,b}$, respectively.
While the rates $k_{X,\a}$, $\a=a, b$ depend on the external force, $k_{X,\a}=k_{X,\a}(\fext)$, the exchange rates $\g_{X;\a}$ are assumed to be independent of $\fext$.
}
\label{Fig9}
\end{figure}

The TCM represents the simplest possible model for dynamic disorder.
This model is defined as consisting of two configurations $a$ and $b$ in each ensemble $A$ and $B$, cf. Fig.\ref{Fig9}.
This kinetic scheme is irreducible in general. However, there are some limits that give rise to reducible schemes.
In the present paper we assume that there is a common equilibrium constant,
\be\label{Rates.TCM}
k_{B,\a}(\fext)=K(\fext)k_{A,\a}(\fext)
\quad \a=a,\,b
\ee
meaning that we are essentially treating a fluctuating barrier model\cite{Brown:2003}.
We additionally assume that the exchange rates $\g_{X;b}$ and $\g_{X;a}$ have the same value,
\be\label{gamma.X.def}
\g_X=\g_{X;a}=\g_{X;b}
\quad;\quad X=A,\,B
\ee
In the following, we will not explicitly indicate the force-dependence of the rates and only write
$k_{X,\a}$ and $K$.
Furthermore, it is reasonable to assume that the exchange rates $\ga$ and $\gb$ show a different dependence on the applied force than the $k_{X,\a}$. 
For simplicity, we assume the that exchange rates do not at all depend on the external force,
$\g_X(\fext)=\g_X(0)$.

Most quantities of interest are determined by the mean rate and the standard deviation for $X=A,B$:
\be\label{kXM.kXD.def}
\kXm=\lg k_X\rg={1\over2}\left(k_{X,a}+k_{X,b}\right)
\quad;\quad
\kXd=\s_X=\sqrt{\lg\d k_X^2\rg}={1\over2}\left(k_{X,a}-k_{X,b}\right)
\ee
Notice that due to eq.(\ref{Rates.TCM}) we have $\kBm=K\kAm$ and $\kBd=K\kAd$.
\subsubsection*{Statistics of transitions}
For the TCM one explicitly has the following expression for the relevant matrix of transition rates.
According to the definition in Fig.\ref{Fig9} and eq.(\ref{gamma.X.def}) one has:
\be\label{W.TCM}
\W=\left(\begin{array}{cccc}
-k_{A,a}-\ga & \ga & k_{B,a} & 0\\
\ga & -k_{A,b}-\ga & 0 & k_{B,b}\\
k_{A,a} & 0 & -k_{B,a}-\gb & \gb\\
0 & k_{A,b} & \gb & -k_{B,b}-\gb
\end{array}\right)
\ee
from which one easily constructs the matrices $\W'$ and $\V$ according the definitions in 
eq.(\ref{V.cycle}) and (\ref{V.events}).
Using the corresponding expressions in the definition of the generating function, eq.(\ref{F.zt}), and the respective derivatives, eq.(\ref{N.Nsquare}), the mean number of transitions and the Q-parameter can be calculated analytically.
The long-time limits of the mean number of transitions are:
\be\label{N.TCM}
\lg N(t)\rg^{(ev)}=2\lg N(t)\rg^{(c)}=2{K\over(1+K)}\kAm t
\ee
In case of cycle-counting, this result has been given earlier\cite{DJ:2009,Brown:2006}.
The force-dependence of this quantity has been discussed in I. 
Also the Mandel parameter $Q(t)$ can be calculated analytically.
The general expressions for the time-dependence of $Q(t)$, however, are not very illustrative.
Starting from $Q(t=0)=0$, the long-time limit
\be\label{Qinfty.TCM}
Q_\infty^{(ev)}={(K-1)^2\over(K+1)^2}+\chi{K\over K+1}
\quad\mbox{and}\quad
Q_\infty^{(c)}=-2{K\over(K+1)^2}+{\chi\over2}{K\over K+1}
\ee
is reached on a time scale that depends strongly on the exchange rate.
Here, the deviations from Markovian behavior are comprised in the function $\chi$, given by
\be\label{chi.TCM}
\chi={2\kAd^2\left[(1+K)^2(\kAm^2-\kAd^2)+2\kAm(\ga+K\gb)\right]
\over
(1+K)\kAm\left[(K\ga+\gb)(\kAm^2-\kAd^2)+2\ga\gb\kAm\right]}
\ee
which in the case of a common exchange constant $\g=\ga=\gb$ reduces to the expression given in I,
$\chi=2\kAd^2/(\g\kAm)$.
For small values of the exchange rate, it takes very long until the limit is reached.
This can be understood from the limiting behavior for the case of static disorder, $\g_X\to0$: 
\be\label{Q.gamto0.TCM}
Q(t)^{(ev)}=2Q(t)^{(c)}={2\kAd^2\over\kAm}{K\over1+K}t
\ee
as already found in I\cite{I.comment}.
\subsubsection*{Waiting time distributions}
For the WTDs the following bi-exponential expressions are obtained from the definition, 
eq.(\ref{Phi.X.def}):
\Be\label{PhiX.TCM}
\Phi_X(t)=
&&\hspace{-0.6cm}
{e^{-(\g_X+\kXm)t}\over\kXm W_X}
\left[
(\kXm^2+\kXd^2)W_X\cosh{(W_Xt)}\right.
\nonumber\\
&&\hspace{1.6cm}+
\left.
\left(\g_X(\kXm^2-\kXd^2)-2\kXm\kXd^2\right)\sinh{(W_Xt)}
\right]
\Ee
with $W_X=\sqrt{\g_X^2+\kXd^2}$.\\
Furthermore, one has:
\Be\label{PhiAB.TCM}
\D\Phi_{A,B}(t_1,t_2)=
&&\hspace{-0.6cm}
\Phi_{A,B}(t_1,t_2)-\Phi_A(t_1)\Phi_B(t_2)
\nonumber\\
=&&\hspace{-0.6cm}
{\kAd^2(\kAm^2-\kAd^2)\over \kAm^2W_{\rm A}W_{\rm B}}K
e^{-(\ga+\kAm)t_1}e^{-(\gb+\kBm)t_2}\times
\nonumber\\
&&\hspace{1.6cm}\times
\left[
(\ga+\kAm)\sinh{(W_{\rm A}t_1)}-W_{\rm A}\cosh{(W_{\rm A}t_1)}\right]\times
\\
&&\hspace{1.6cm}\times
\left[
(\gb+\kBm)\sinh{(W_{\rm B}t_2)}-W_{\rm B}\cosh{(W_{\rm B}t_2)}\right]
\nonumber
\Ee
which is symmetric, $\D\Phi_{A,B}(t_1,t_2)=\D\Phi_{B,A}(t_2,t_1)$.

\noindent
The moments of $\Phi_X(t)$ are calculated according to eq.(\ref{Moms.PhiX}) and are given by:
\be\label{tau.n.x.TCM}
\lg\t\rg_X={1\over\kXm}\quad;\quad
\lg\t^2\rg_X=\left({2\over\kXm}\right){2\g_X+\kXm\over\kXm^2-\kXd^2+2\g_X\kXm}
\ee
Furthermore, we give the moments of the joint distribution $\Phi_{X,Y}(t_1,t_2)$ in the form of the 'deviations from reducibility':
\be\label{del.tau.n.XY.def}
\lg\d\t^n\rg_{XY}=\lg\t^n\rg_{XY}-\lg\t^n\rg_X\lg\t^n\rg_Y
\ee
For the TCM these moments are symmetric, $\lg\d\t^n\rg_{XY}=\lg\d\t^n\rg_{YX}$.\\
With the abbreviation
\be\label{Nenn.tau}
N_{XY}:=\left(\kAm^2-\kAd^2+2\ga\kAm\right)\left(K(\kAm^2-\kAd^2)+2\gb\kAm\right)
\ee
one finds:
\be\label{del.tau.n.XY.res.einK}
\lg\d\t\rg_{AB}={\kAd^2\over \kAm^2}{\kAm^2-\kAd^2\over N_{XY}}
\quad\mbox{and}\quad
\lg\d\t^2\rg_{AB}={16(\kAm^2-\kAd^2)\kAd^2\over K\kAm^2}
{(\ga+\kAm)(\gb+\kBm)\over N_{XY}^2}
\ee
\end{appendix}

\begin{thebibliography}{10}
\bibitem{I.comment}
note that in I there is a factor $K/(1+K)$ missing in the case of static
  disorder.
\end{thebibliography}
\end{document}